\documentclass{article}
\usepackage[utf8]{inputenc}
\pdfpagewidth\paperwidth
\pdfpageheight\paperheight
\usepackage[T1]{fontenc}
\usepackage[utf8]{inputenc}
\usepackage[english]{babel}
\usepackage{amsmath}
\usepackage{amsfonts}
\usepackage{graphicx}
\usepackage{subcaption}
\usepackage{amsthm}
\usepackage{graphicx}
\usepackage{amssymb}
\usepackage{geometry}
\usepackage{subcaption}
\usepackage{caption}
\usepackage{float}
\usepackage{bm}
\usepackage{titling}

\usepackage[table]{xcolor}
\usepackage[title]{appendix}
\usepackage{braket}

\usepackage[backend=bibtex,sorting=none]{biblatex}
\def\qe{\textsc{Quantum ESPRESSO}\texttrademark \ }

 \title{Evidence of ferroelectric features in low-density supercooled water from \textit{ab initio} deep neural-network simulations}
\author{Cesare Malosso$^1$ \and Natalia Manko$^2$ \and Maria Grazia Izzo$^1$ \and Stefano Baroni$^1$ $^4$ \and Ali Hassanali$^2$}

\date{
	$^1$SISSA -- Scuola Internazionale Superiore di Studi Avanzati, Trieste (Italy) \\ %
	$^2$The Abdus Salam Centre for Theoretical Physics, Trieste (Italy)\\
        $^3$ CNR-IOM, Istituto dell'Officina dei Materiali, SISSA unit, Trieste (Italy)\\[2ex]%
}
\begin{document}
\maketitle
\begin{abstract}
Over the last decade, 
an increasing body of evidence has emerged, supporting
the existence of a metastable liquid-liquid critical point in supercooled water, whereby two distinct liquid phases of different densities coexist. 
Analysing long molecular dynamics simulations performed using deep neural-network force fields trained to accurate quantum mechanical data, 
we demonstrate that the low-density liquid phase displays a strong propensity toward spontaneous polarization, as witnessed by large and long-lived collective dipole fluctuations. Our findings suggest that the dynamical stability of the low-density phase, and hence the transition from high-density to low-density liquid, is triggered by a collective process involving an accumulation of rotational angular jumps, which could ignite large dipole fluctuations. This dynamical transition involves subtle changes in the electronic polarizability of water molecules which affects their rotational mobility within the two phases. These findings hold the potential for catalyzing new activity in the search for dielectric-based probes of the putative second critical point.
	\end{abstract} \hspace{10pt}

Water is a universal solvent that sustains the thermodynamics and dynamics of numerous physical, chemical and biological processes \cite{Ball2007}. Unlike simple liquids, water is characterized by several anomalies that are manifested in different regions of the phase diagram. These include, for example, the presence of the density maximum at 4$^{\circ}$C leading to a lower density of the solid phase of water (ice) compared to the liquid phase, as well as the increase in compressibility of liquid water under supercooling, to name a few \cite{Speedy1976,Mishima1998,Debenedetti2003}. Despite significant efforts from experimental, theoretical, and computational fronts, consensus on the microscopic origins of water's anomalies has not yet been achieved, thus nurturing an open and active area of research \cite{Nilsson2015,Russo2018p,Shi2020}.

Currently one of the most lively areas of study in the chemical physics of water pertains to the existence of a second critical point in deeply supercooled water. Several decades ago, Poole and co-workers put forward a fascinating suggestion \cite{Poole1992} that liquid water may undergo a first-order liquid-liquid transition (LLT), resulting in the coexistence of two metastable, high- and low-density, liquid phases, hereafter referred to as HDL and LDL, respectively. The existence of this transition is thought to possibly rationalize many of liquid water's peculiar properties \cite{Russo2014,Russo2018}. The experimental proof of the existence of a second critical point is extremely challenging owing to the rapid crystallization under supercooled conditions. Recent experiments, however, appear to break the boundary of no-man's land and give stronger evidence for the existence of the liquid-liquid critical point (LLCP) \cite{AmannWinkel2013,Kim2020,AmannWinkel2023}.

Due to the experimental challenges in probing the thermodynamics in the supercooled regime, numerical simulations continue to play a crucial role in understanding the origins of the phenomena. In particular, the ST2 water model has been unequivocally shown to exhibit an LLCP \cite{Liu2009,Debenedetti2020,Kesselring2012,Poole2013,Palmer2014}. In addition a few years ago, microsecond long simulations of one of the most realistic classical models of liquid water, TIP4P/2005 \cite{tip4p}, have confirmed the existence of the LLCP \cite{Debenedetti2020}. Advancements in machine-learning potentials have also opened up the possibility for highly accurate predictions with only a marginal increase in computational cost compared to classical force fields. In this context, Behler and Parrinello pioneered the development and application of the first neural-network (NN) potentials \cite{Behler2007} that were later used to study both thermodynamics and dynamics of water across the phase diagram \cite{Morawietz2012,Morawietz2013,Morawietz2016,Cheng2019}.
 
Over the last decade, Car and co-workers have expanded the scope of the original NNs with the development of deep neural-network (DNN) potentials \cite{Weinan2018,Wang2018,Zhang2018,Zhang2020}. The infrastructure provided by DNNs includes highly robust algorithmic advances within an open-source software allowing for an automatic generation and training of \textit{ab inito} quality potentials \cite{deepmdv2,Zhang2020_1}. These advancements have paved the path for a systematic exploration of bulk thermodynamic properties \cite{Jiang2021,Zhang2021,Zhang2021_1,Piaggi2021,Piaggi2022}, as well as dynamic and transport properties \cite{Li2020,Tisi2021,Malosso2022} that were previously inaccessible through direct DFT methods, or with many-body polarizable potentials such as MB-pol \cite{Reddy2016,Babin2013,Babin2014,Medders2014}.

More recently, Gartner and Car have deployed these DNN potentials to study the behavior of liquid water upon supercooling and also find evidence for the existence of a LLCP \cite{Gartner2020,Gartner2022}. Given that these DNN simulations come from very high quality electronic structure calculations, namely, using the accurate meta-GGA exchange-correlation functional (SCAN) \cite{SCANPerdew,Sun2016}, further strengthens the hypothesis that water can separate into two distinct liquids under supercooled conditions. 

In this work, we use these DNN trajectories that were made publicly available in Ref. \cite{Gartner2022}, to examine the coupling between fluctuations in density and polarization. By designing an additional deep neural-network that is trained to reproduce the electronic polarizability \cite{Zhang2020,Sommers2020} in the system, our analysis of the 0.5 microsecond molecular dynamics trajectories reveal that the LDL phase displays a tendency to exhibit ferroelectric character as revealed by the presence of spontaneous polarization in the liquid. The dynamical stability of this polarization in the LDL liquid is found to be strongly coupled to the concentration of orientational defects in the hydrogen-bond network. We show that the transitions between the LDL and HDL phase are activated by a collective reorientational process of large angular jumps \cite{Laage2006} and offer new perspectives on how these dielectric properties may serve as another probe of the long-sought critical point.

\section*{Results}
\subsection*{Polarization fluctuations}
\begin{figure*}[t]
\captionsetup[subfigure]{labelformat=empty}
\begin{subfigure}{0.5\linewidth}
\centering
\subfloat[]{\includegraphics[width=\linewidth]{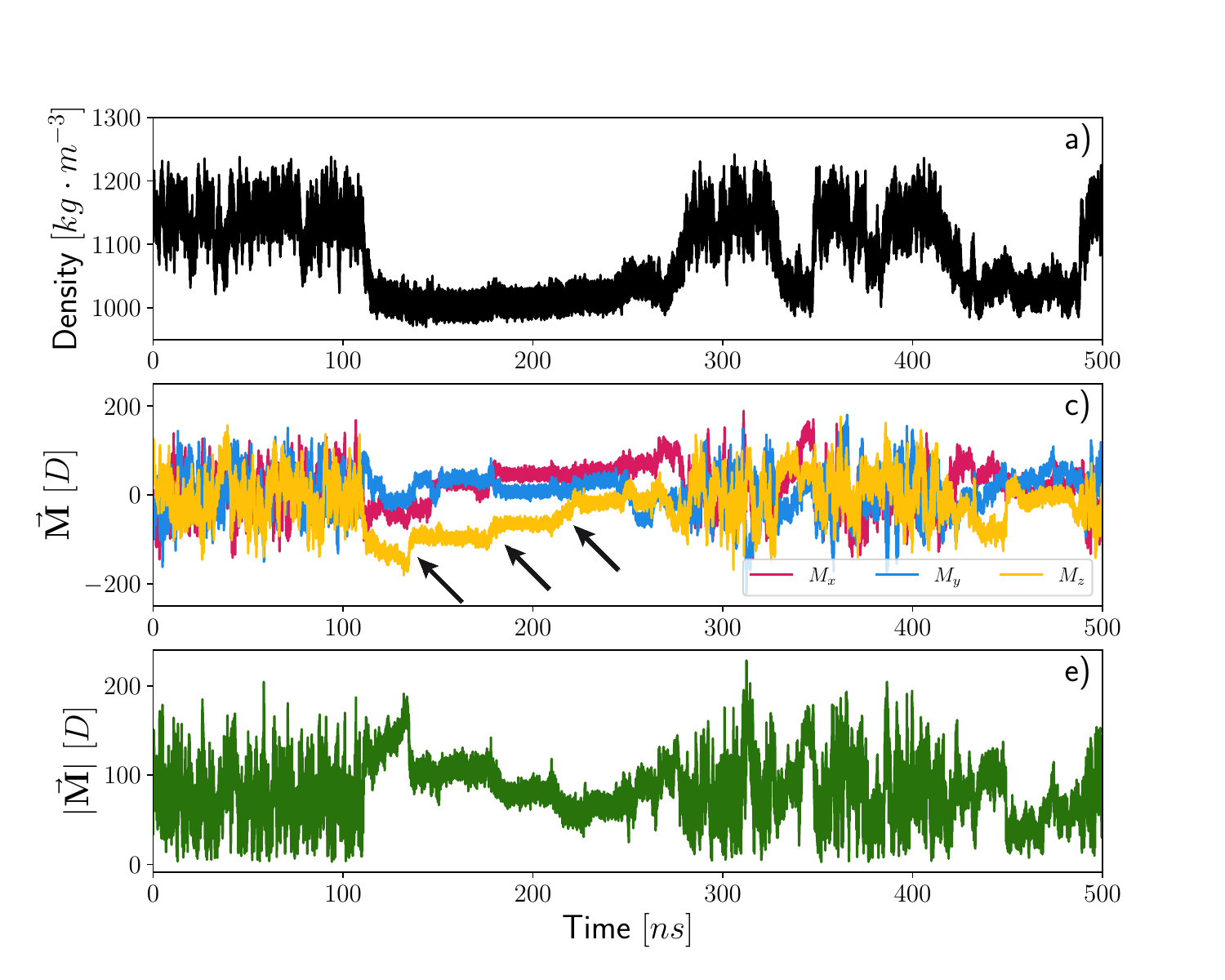}}
\end{subfigure}%
\begin{subfigure}{0.5\linewidth}
\centering
\subfloat[]{\includegraphics[width=\linewidth]{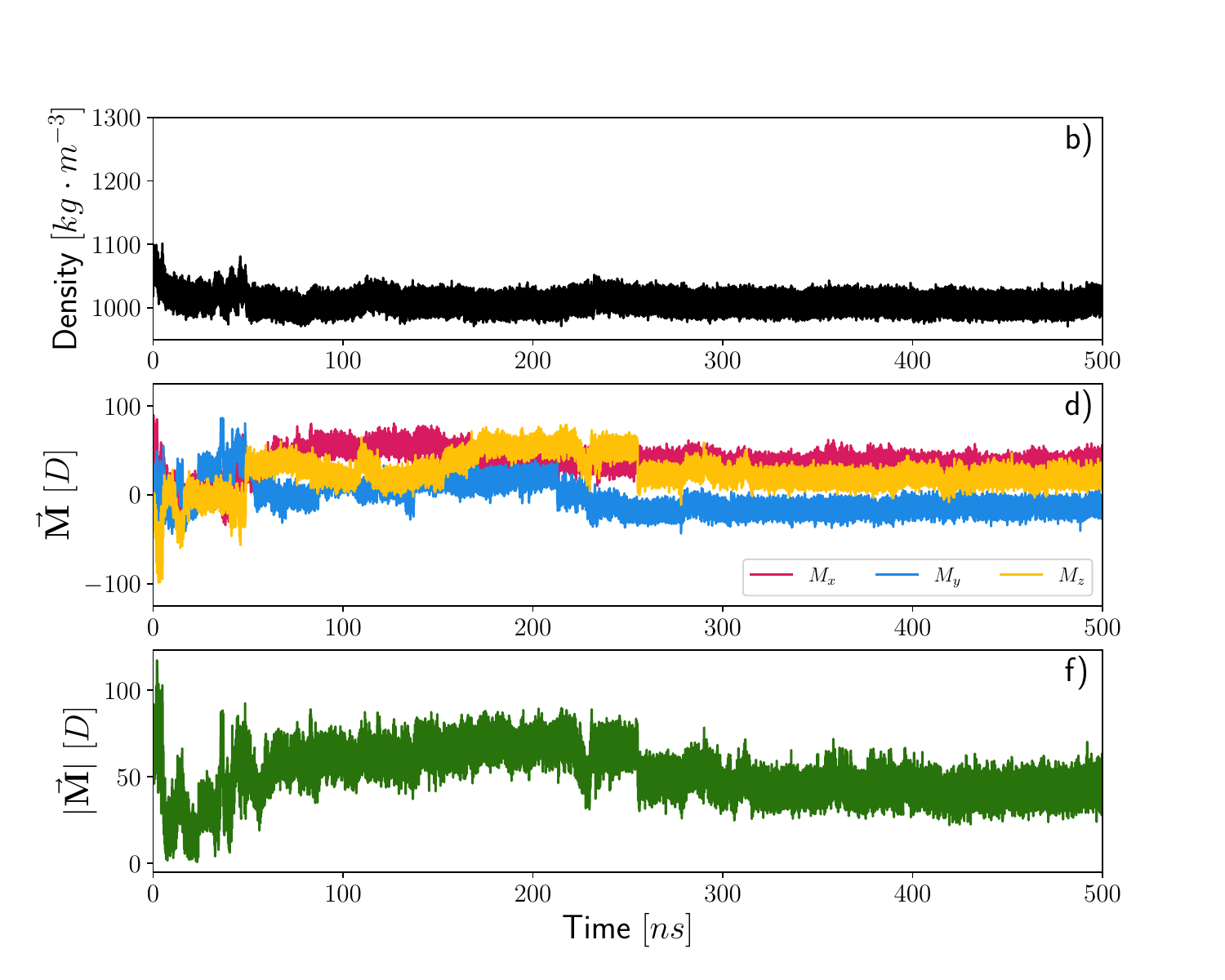}}
\end{subfigure}%
\caption{Density[\textit{top}], total dipole[\textit{middle}] and its modulus[\textit{bottom}] fluctuations during isobaric-isothermal molecular dynamics simulations at nearly supercritical conditions - 235K and 3200bar - for two different \textit{ab initio} SCAN DFT simulations. The pronounced variations in density offer compelling proof of the presence of a metastable critical point. Both simulations consistently display spontaneous polarization in the low-density phase. This tendency is revealed by the long-standing nearly constant values of the total dipole in the LDL phase as well as a symmetry breaking of the dipole components.}
     \label{fig:dipole}
 \end{figure*}
 
Leveraging insights from prior studies, we turn to examining DNN molecular dynamics (MD) simulations of supercooled water reported by Gartner \textit{et al.}. Remarkably, this model also displays evidence for a second critical point at low temperatures and high pressures, giving rise to a low-density and a high-density liquid phase, in line with other classical water models. Since the LDL phase is thought to be characterized by local order akin to what one would expect in ice, we hypothesized that the HDL-LDL transition could be reflected in changes in its orientational order and possibly be manifested in the polarization of the system.

The macroscopic polarization of an extended system of charged particles is ill-defined in periodic boundary conditions (such as used in any MD simulation), unless each charge can be unequivocally assigned to a neutral, non-dissociating unit \cite{Resta_Vanderbilt_2007}. 
As long as the molecules do not dissociate, their dipole can be estimated by associating to each of them a well defined number of Wannier centers, computed from Maximally Localized Wannier Functions (MLWF) \cite{wannier,Souza2001}. Therefore,
we crafted a separate deep neural-network \cite{Zhang2020} trained to predict the dipole of each water molecule of the system from SCAN-DFT data. This dipole DNN is used to predict, starting from the atomic positions, the molecular dipoles from which the macroscopic polarization can trivially be obtained.

The results of this analysis are summarized in Fig.\ref{fig:dipole} for the two different isobaric-isothermal simulations reported by Gartner \textit{et al.}. The different subplots show the time series of the density of the system (a)-b)), each of the components of the total dipole (c)-d)), and finally, the dipole modulus (e)-f)). The prominent density fluctuations provide clear evidence of the existence of a metastable critical point near the pT conditions of the corresponding simulations (235K and 3200bar). Specifically, Fig. \ref{fig:dipole} a) is an example of a trajectory where we observe $\sim$5 transitions between the HD and LD phases. On the other hand, in panel b) the system remains in the LD phase for the entire length of the 500-ns simulation.

Turning next to the dielectric properties of supercooled water undergoing the LLT, the time series of the total dipole $\bf{M}$ (depicted in Fig. \ref{fig:dipole} c) and d)) reveals a strong inclination of the low-density phase towards spontaneous polarization. In particular, LDL water exhibits rather long-lived and nearly constant values of the dipole components surviving for timescales of periods of approximately 100ns. Interestingly, the extent of the polarization both in terms of the magnitude and direction displays a breaking of symmetry. The bottom most panel shows the total dipole moment in the system as a function of time where one clearly sees a marked change in the magnitude of the fluctuations between the LDL and HDL. Similar features are also seen in both the trajectory that hops between the two phases (panel a) and the other which is trapped in the LD phase (panel b). In contrast, in the HDL, the dipole components are symmetrically centered around zero as one would expect in a more disordered phase.   

\begin{figure*}[t]
    \centering
    \includegraphics[width=7cm]{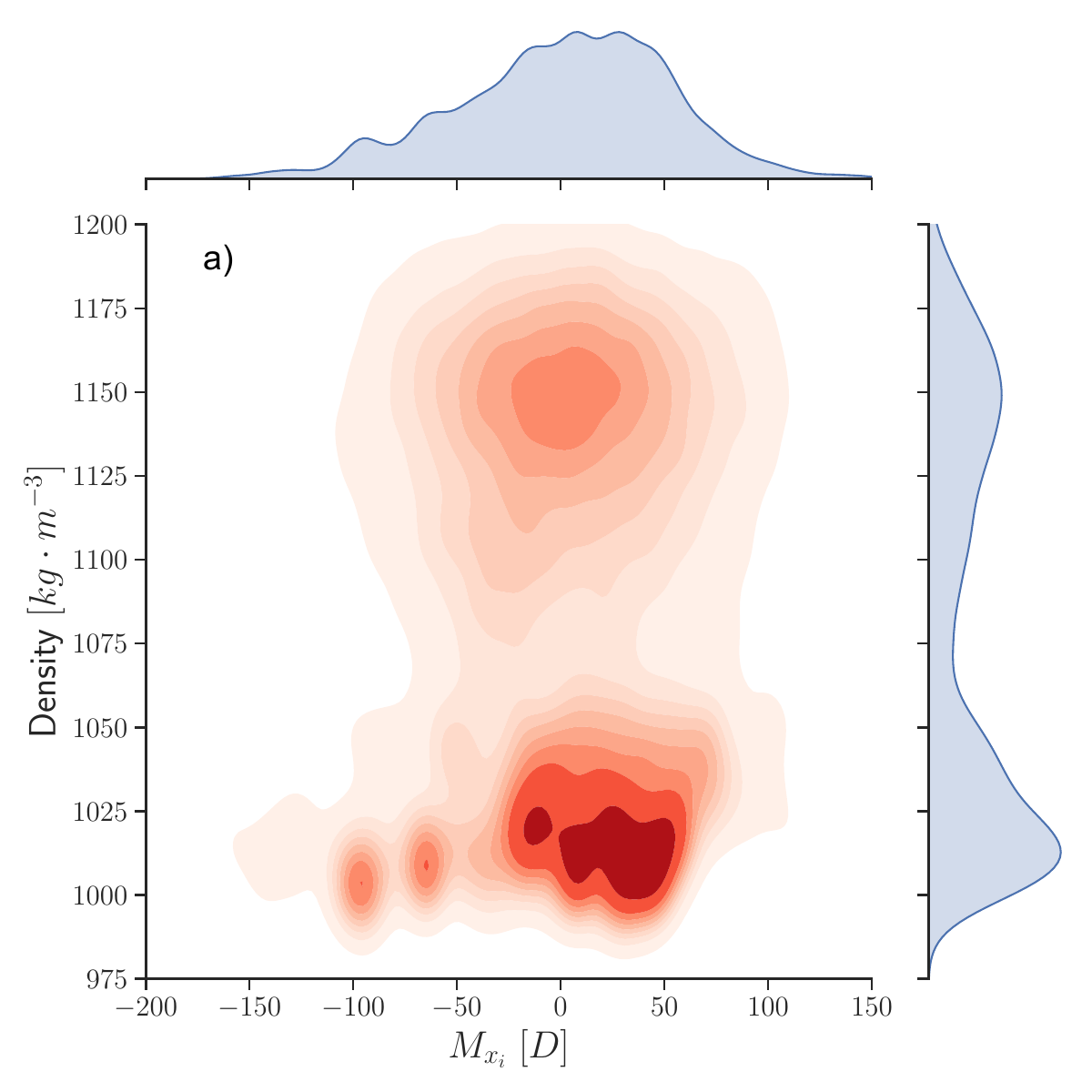}
    \includegraphics[width=7cm]{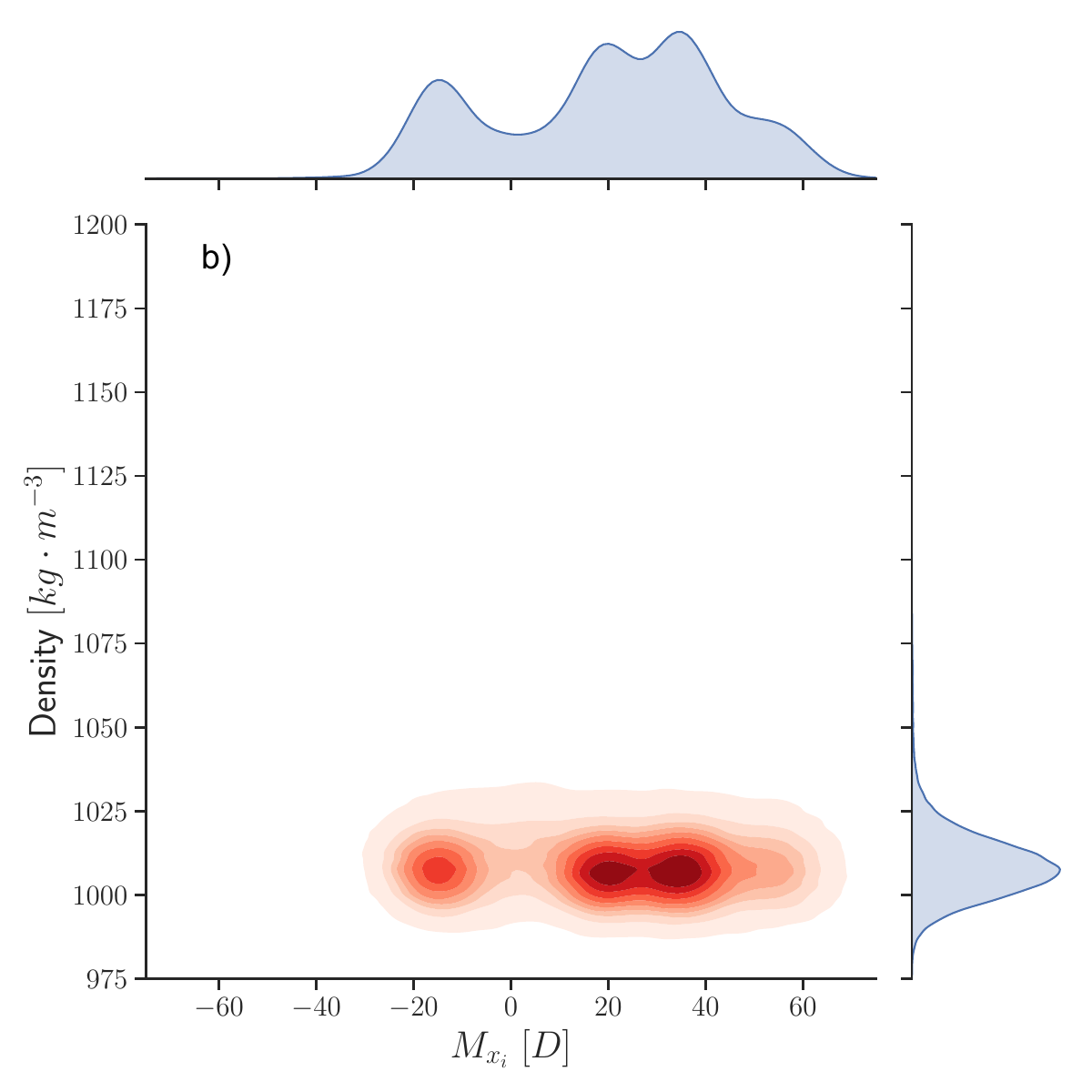}
    \caption{Statistics of critical fluctuations displayed as two-dimensional density and dipole probability densities for the SCAN-DFT simulations shown earlier. The polarization within the HDL phase is centered around zero while in the LDL phase, there are some underlying structure.}
    \label{fig:kde}
\end{figure*}

The preceding results strongly point to the importance of the polarization order parameter in affecting the kinetics of the LDL to HDL transition. While the correlation time needed to converge the dipole fluctuations in the LDL phase clearly require significantly longer trajectories that is beyond the scope of the current work and methods employed, it is instructive to construct probability distributions along the density and polarization coordinates. Figure \ref{fig:kde} a)-b) shows the two-dimensional distributions for the mass density and dipole (including the three components) as well as the marginal distributions. These density plots firstly confirm that the HD-LD transition is essentially dominated by a mode moving along the density coordinate (y-axes). On the other hand, looking along the x-axis one observes significant structure in the dipole coordinate. Specifically, in both the trajectories (left and right panel), there are at least three different minima that appear. In the case of the simulation that remains trapped in the LDL phase, the marginal distribution along the dipole, clearly shows strong evidence for polarization on the timescales of the current simulation.

\subsection*{Angular Jumps and the LD-HD Transition}
\renewcommand\thesubfigure{\arabic{subfigure}}
\begin{figure*}[t]
\captionsetup[subfigure]{labelformat=empty}
\begin{subfigure}{0.5\linewidth}
\centering
\subfloat[]{\includegraphics[width=\linewidth]{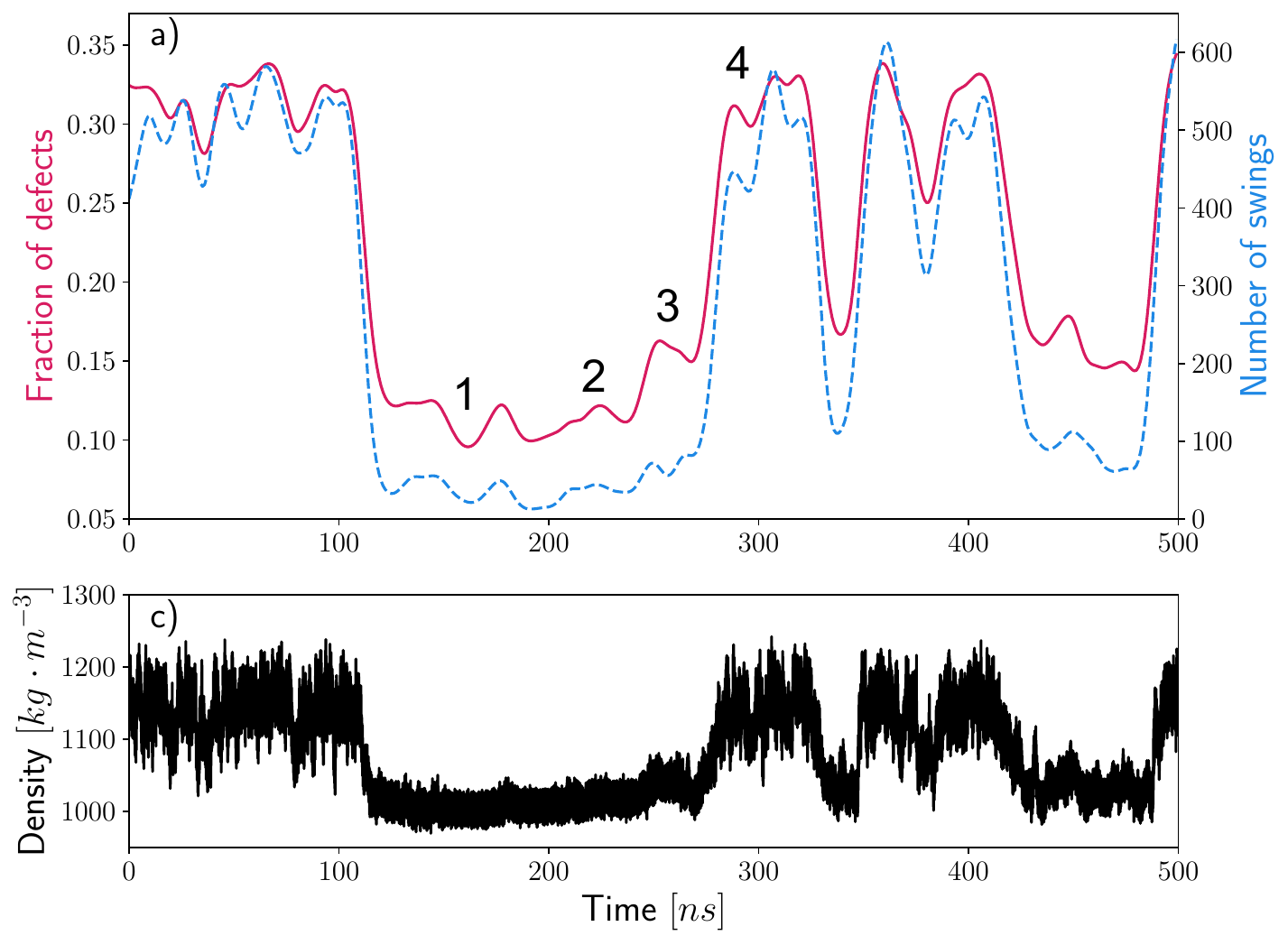}}
\end{subfigure}%
\begin{subfigure}{0.5\linewidth}
\centering
\subfloat[]{\includegraphics[width=\linewidth]{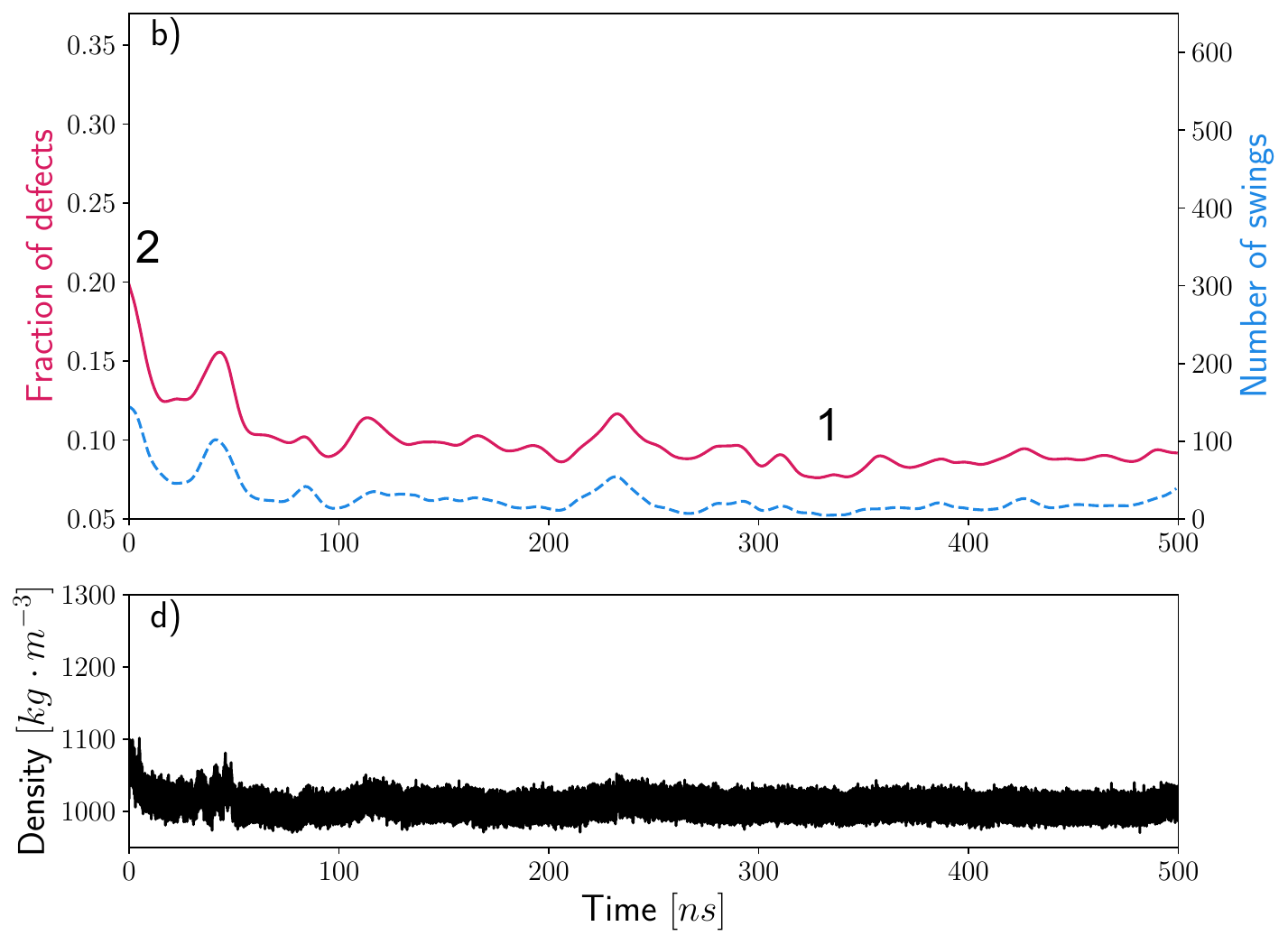}}
\end{subfigure}\par\medskip
\centering

\captionsetup[subfigure]{font=Large,labelfont=Large,labelformat=simple}
\setcounter{subfigure}{0}
\begin{subfigure}{0.20\linewidth}
\centering
\subfloat[]{\includegraphics[width=\linewidth]{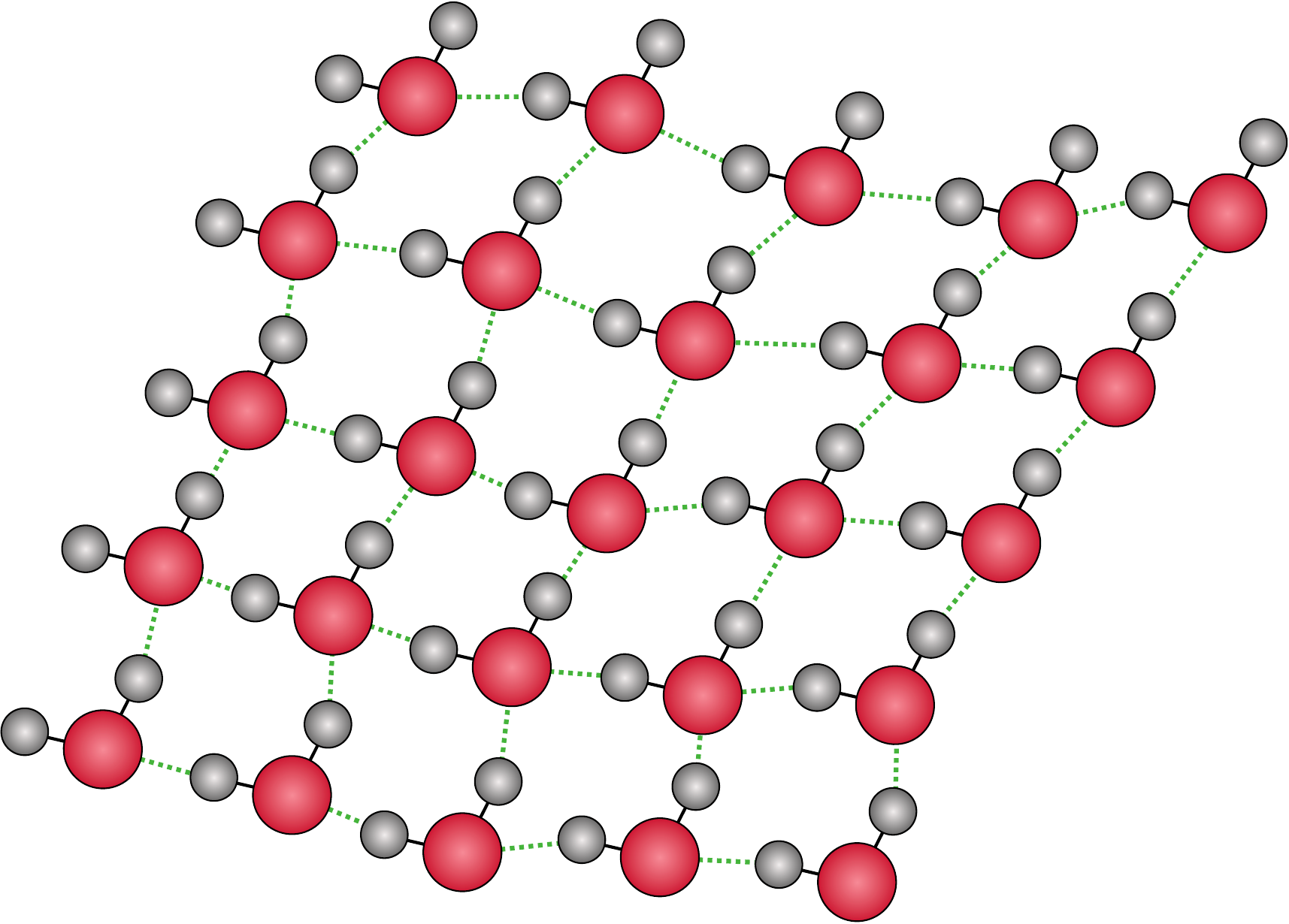}}
\end{subfigure}%
\begin{subfigure}{0.20\linewidth}
\centering
\subfloat[]{\includegraphics[width=\linewidth]{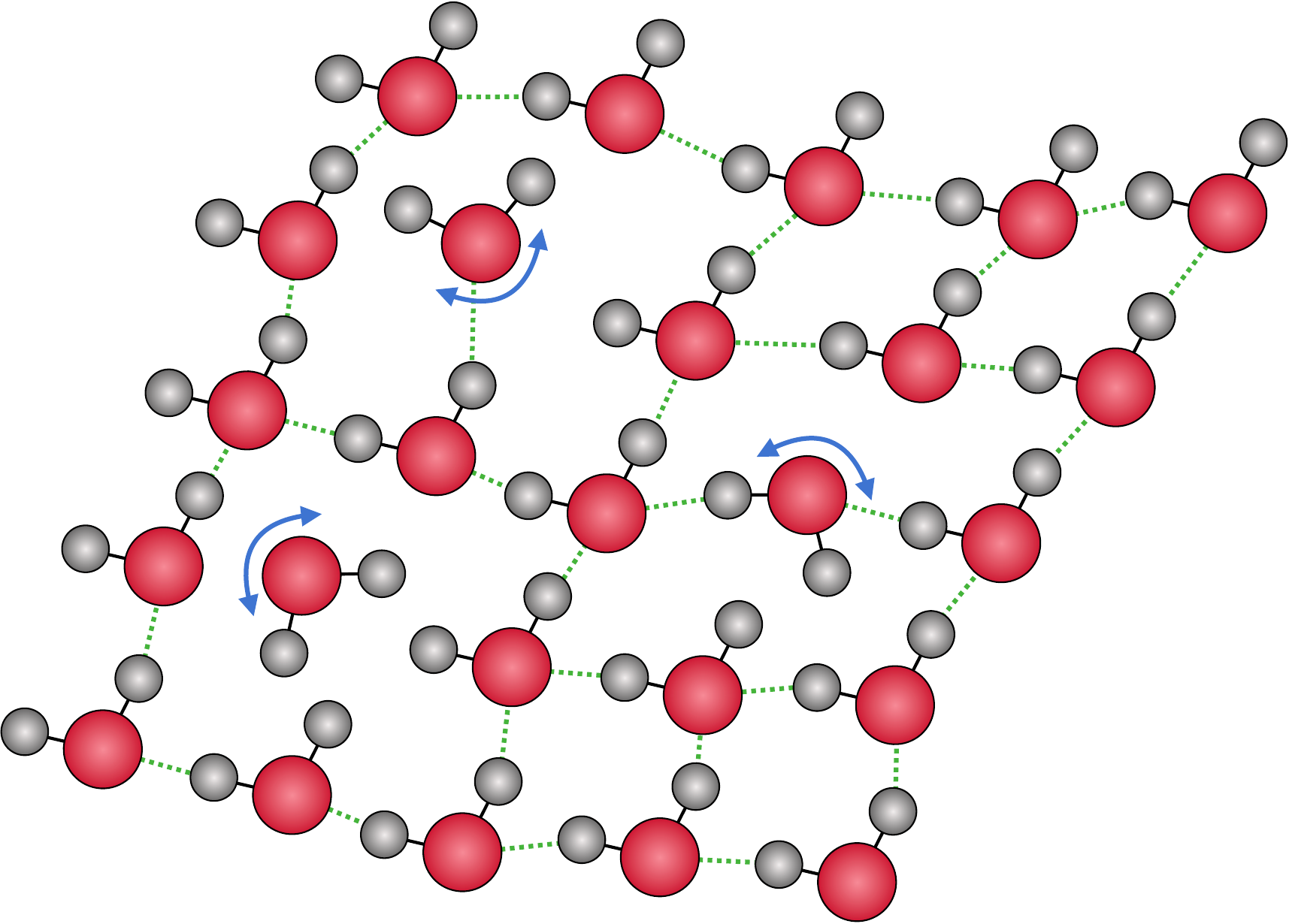}}
\end{subfigure}
\begin{subfigure}{0.20\linewidth}
\centering
\subfloat[]{\includegraphics[width=\linewidth]{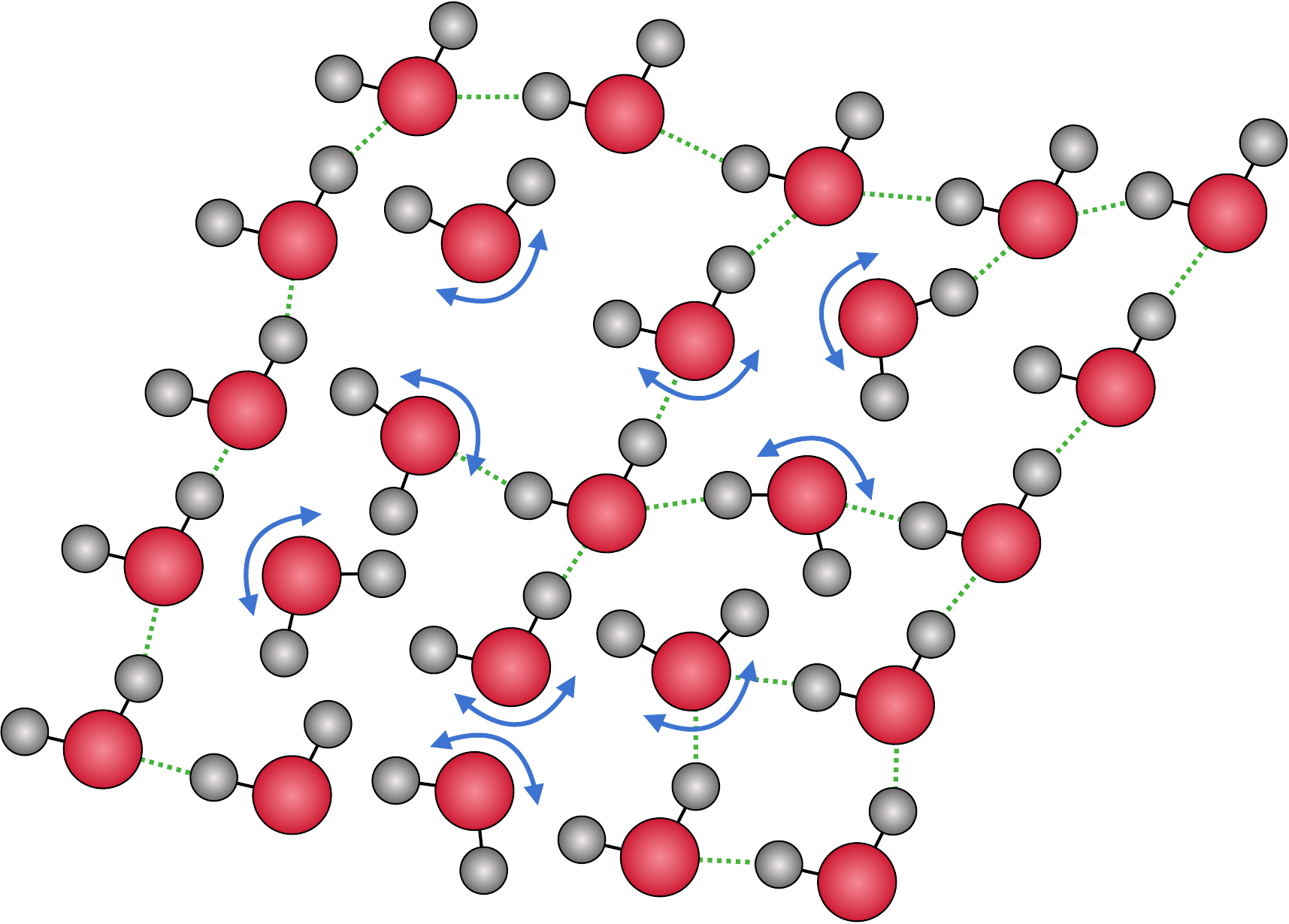}}
\end{subfigure}%
\begin{subfigure}{0.20\linewidth}
\centering
\subfloat[]{\includegraphics[width=\linewidth]{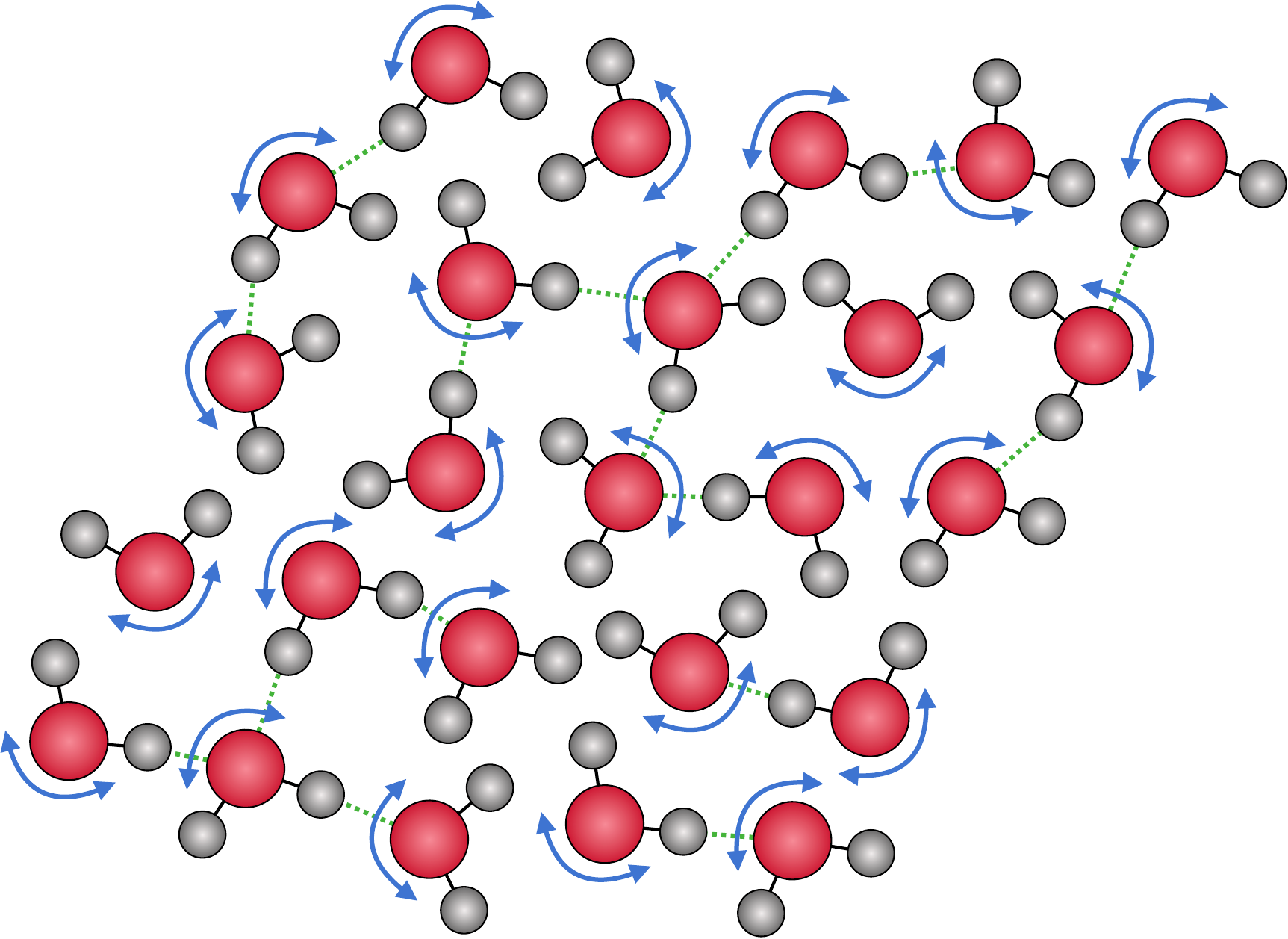}}
\end{subfigure}
\caption{Panels a and b illustrate the number of angular jumps (right axes) and defects (left axes) obtained by analyzing the two trajectories close to the critical point. The middle panels c and d instead reproduce the density time-series as a guide to the readers eye to compare with the reorientational dynamics that emerges from the top row. Finally, the bottom panel serves as as visual schematic of the underlying molecular mechanism that couples the collective reorientational motions of water molecules and changes in polarization.}
\label{fig:jumps}
\end{figure*}

A careful examination of the time series associated with the dipole shows the presence of kinks where there are relatively large and abrupt changes in the dipole values (see arrows in the Figure that point to these features). The presence of the kinks is more enhanced in the left panel owing to the fact that 
in this trajectory, the system oscillates between the LDL and HDL phase and thus the former is metastable. In the second trajectory, the dipole assumes nearly constant values for hundreds of nanoseconds with much fewer kinks.

Changes in the macroscopic dipole and its components over time strongly suggests that there are likely some important changes in the orientational correlations of the hydrogen-bond network network during the LDL-HDL transitions. At ambient temperature, rotational dynamics in liquid water has been shown to involve large angular jumps \cite{Laage2006} which lead to the creation of coordination defects. These types of coordination defects also play an important role in different regions of the phase diagram. In ice for example, the migration of orientational Bjerrum-defects involves a highly cooperative reorganization of the network\cite{Bjerrum1952,deKoning2020}. Similarly, previous studies by Sciortino have demonstrated that coordination defects in water serve as catalytic sites for facilitating the motion of water molecules \cite{Sciortino1991,Sciortino1996}. 

Recently, some of us have shown that angular jumps in liquid water involves a highly cooperative process involving the fluctuations of many hydrogen-bond defects in the network \cite{OffeiDanso2023}. We hypothesized that the kinks observed in our dipole time series maybe coupled to these types of collective reorientational dynamics. We thus applied an automatized protocol that searches the time-series of all dipole components of each water molecule identifying regions where there are sharp changes indicating large angular jumps (see Methods for more details). The two \textit{top} panels show the time evolution of the average number of angular jumps observed over time windows of 1ps for the two trajectories. In addition, we also show the fraction of defects in the network defined by the number of water molecules that break ice-rules of donating and accepting two hydrogen bonds. This was determined by using a geometrical criterion of hydrogen bonds determined in similar spirit to that proposed by Luzar and Chandler \cite{Luzar1996} (see Methods for more details). On the bottom of each panel, the corresponding density time series (\textit{black}) is shown. 

Interestingly, the two phases exhibit markedly different dynamical behavior. The low-density phase appears more structured, with fewer defects and a more glassy nature, constraining rotational mobility. On the other hand, the high-density phase resembles more what one expects in a liquid, showcasing a greater number of defects and increased molecular mobility. Looking specifically at the time interval between 100-300ns where the system resides in the LDL phase, we observe a much smaller number of angular rotations. In this particular case, large jumps are defined by those where the angle change is greater than 60 degrees. More interestingly, the oscillations in the number of angular jumps appear to be rather well correlated with regions in Fig. \ref{fig:dipole} where the kinks in the dipole are observed. The enhancement in the oscillations together with the reduction of orientational order is depicted in the schematic for example going from regions 1 and 2. 

Moving further along that time interval between 100-300ns, the transition from the LDL to the HDL phase involves a growth in the number of rotational jumps (region 3) which ultimately then pushes the system fully into the disordered HDL liquid phase (3-4 in the schematic). Between 300-500ns the system oscillates several times between the HDL and LDL phase as seen in the density plots Fig.\ref{fig:jumps}. In this regime, we observe rather curiously, that the LDL phase appears to be more dynamically unstable yielding period where there are a larger number of rotational jumps occurring compared to regions 1 and 2. These features also appear to be rather well correlated with the presence of kinks in the dipole time series. These dynamical features are also fully consistent with the larger structuring that is observed within the LDL phase in the density-dipole density distributions.

The second trajectory close to the LLCP which essentially remains within the LDL phase, displays very different behavior in the re-orientational dynamics. Here we observe the initial dipole fluctuations and kinks shown earlier are manifested in small magnitude oscillations in the angular jumps. This is also reflected in the decay of the number of defects which is schematically represented by a transition from 2 to 1.

\subsection*{Comparing SCAN-DFT and TIP4P/2005}

\begin{figure}[t]
    \centering
    \includegraphics[width=8cm]{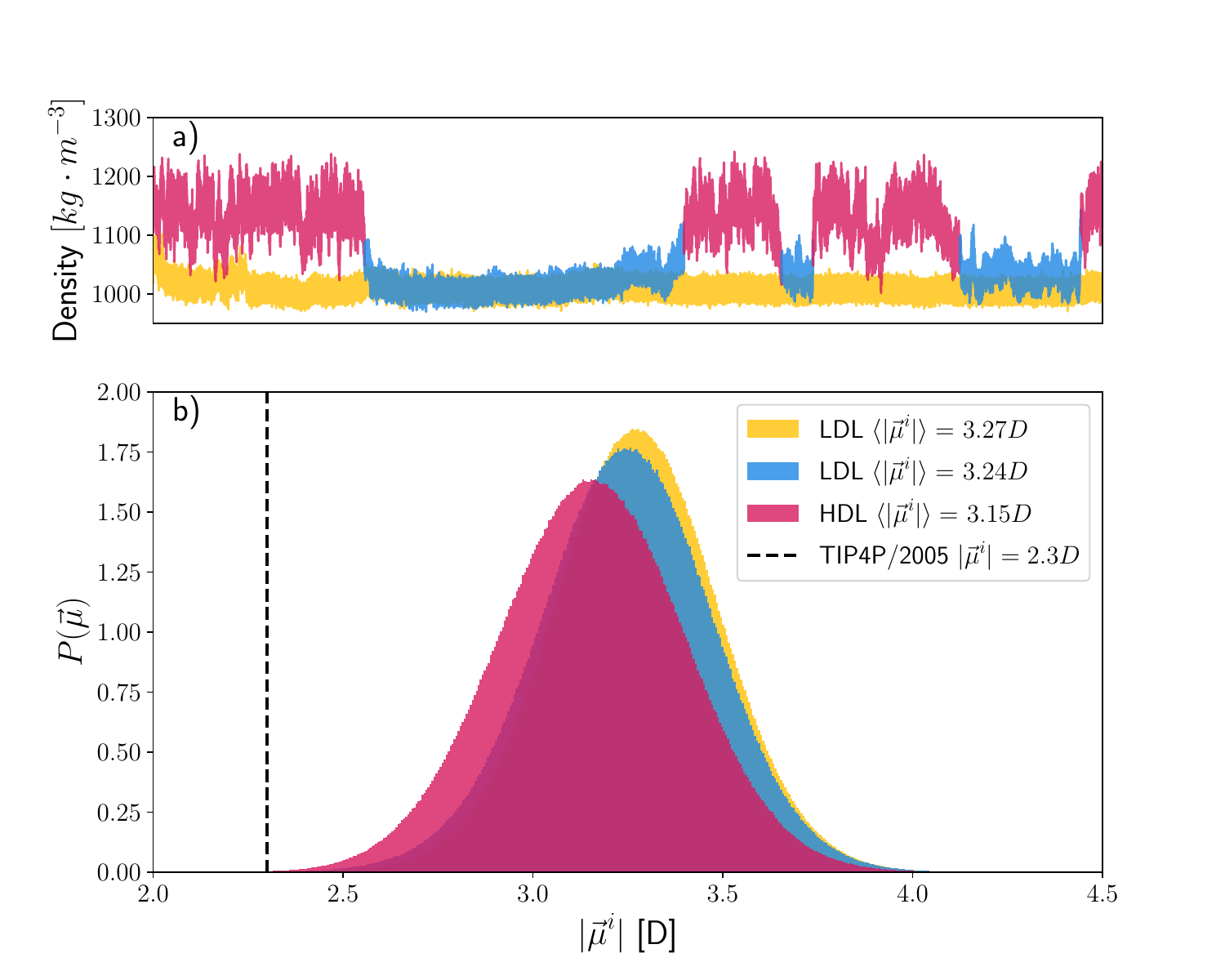}
    \caption{Panel a shows the density time series from two distinct DNN SCAN-DFT simulations, with various phases of supercooled water highlighted in different colors. Panel b depicts the distribution of molecular dipoles for these phases, matching the colors used in panel a. In addition, the value of the molecular dipole as built in the TIP4P/2005 water model is identified by the dashed black vertical line.}
    \label{fig:distr}
\end{figure}

As indicated earlier, Debenedetti and Sciortino demonstrated that TIP4P/2005, one of the most accurate empirical potentials for water, also displays a liquid-liquid critical point at deeply supercooled conditions \cite{Debenedetti2020}. These potentials are however, non-polarizable and therefore can only faithfully account for orientational polarization. It is thus interesting to ask what aspects if any, of the LLT could be altered by explicitly including electronic polarization that is present in the DNN.

Similar to the analysis shown in Fig. \ref{fig:dipole} and Fig. \ref{fig:kde}, we examined the coupling between the density and dipole. As in the DNN simulations, we observe long periods in the LDL phase where there is a tendency for spontaneous polarization. This is seen both in the density distributions and the dipole kinks (see Figure 1a SI). Thus at a qualitative level, the presence of polarization within the LDL phase appears be a generic feature that is independent of the quality of the underlying potential. On the other hand, it is clear that rigid point-charge non-polarizable water models have challenges with making quantitative predictions for dielectric properties\cite{liang2023}. In addition, several studies have shown that point-charge models introduce bigger asymmetries in the electrostatic potential on the accepting versus donating side of hydrogen bonds, compared to electronic structure approaches \cite{Remsing2014}. These differences are also manifested in water dynamics both in terms of its translational diffusion and reorientation polarization \cite{rick_effects_2023}.

One clear distinction between the two potentials is that in TIP4P/2005, the molecular dipole moment is fixed while in the DNN the water molecular are polarizable which will effectively lead to induced dipoles.  In Fig. \ref{fig:distr} we report the distribution of the molecular dipoles computed from the SCAN-DFT trajectories. Different colors are chosen to distinguish the distribution of low- (\textit{magenta}) and high-density (\textit{blue}) water. A third color (\textit{yellow}) is chosen to depict the data taken from the second SCAN-DFT trajectory, where the system is deeply trapped in the low-density phase. Interestingly, we observe rather subtle shifts in the average molecular dipole moment of the LDL vs the HDL regions with the former reflecting an enhancement of the average dipole by 0.12 Debye. However, the fluctuations are rather large with all systems displaying significant overlap. In contrast, the molecular dipole of TIP4/2005 is fixed to 2.3 Debye.

It is beyond the scope of the current contribution to provide a comprehensive analysis of how exactly these differences will manifest in the underlying dynamics associated with the HD-LD transition. We thus attempted to identify some key ingredients associated with the polarization dynamics that we believe warrants further investigation for future studies. Specifically, the autocorrelation of the dipole coming from SCAN appears to display faster relaxation dynamics compared to that of TIP4P/2005 for both the HD and LD phase (see Figure 2 SI). 
The LDL phase exhibits very slow dynamics in the dipole fluctuations and therefore we cannot make any quantitative predictions on its relaxation timescales. For the HDL phase, the relaxation times associated with the dipole for the SCAN and TIP4P/2005 model are ~2ns and ~150ns respectively. Note that the difference in relaxation times between the two models can easily arise from the exponential dependence on small differences in activation barriers between the two models, possibly due to the rigidity of the classical models. In addition, we also examined the relaxation dynamics of the density for the SCAN model. The results presented in Figure 3 in the SI show that for the HDL phase, density and polarization relax on similar timescales, yielding a slow component of approximately 2ns. Conversely, for the LDL phase, it is evident that the dynamics involve fluctuations in both density and polarization, with the latter being significantly slower.
\subsection*{Finite Size Effects on Polarization
}
\begin{figure*}[t]
\captionsetup[subfigure]{labelformat=empty}
\centering
\subfloat[]{\includegraphics[width=\linewidth]
{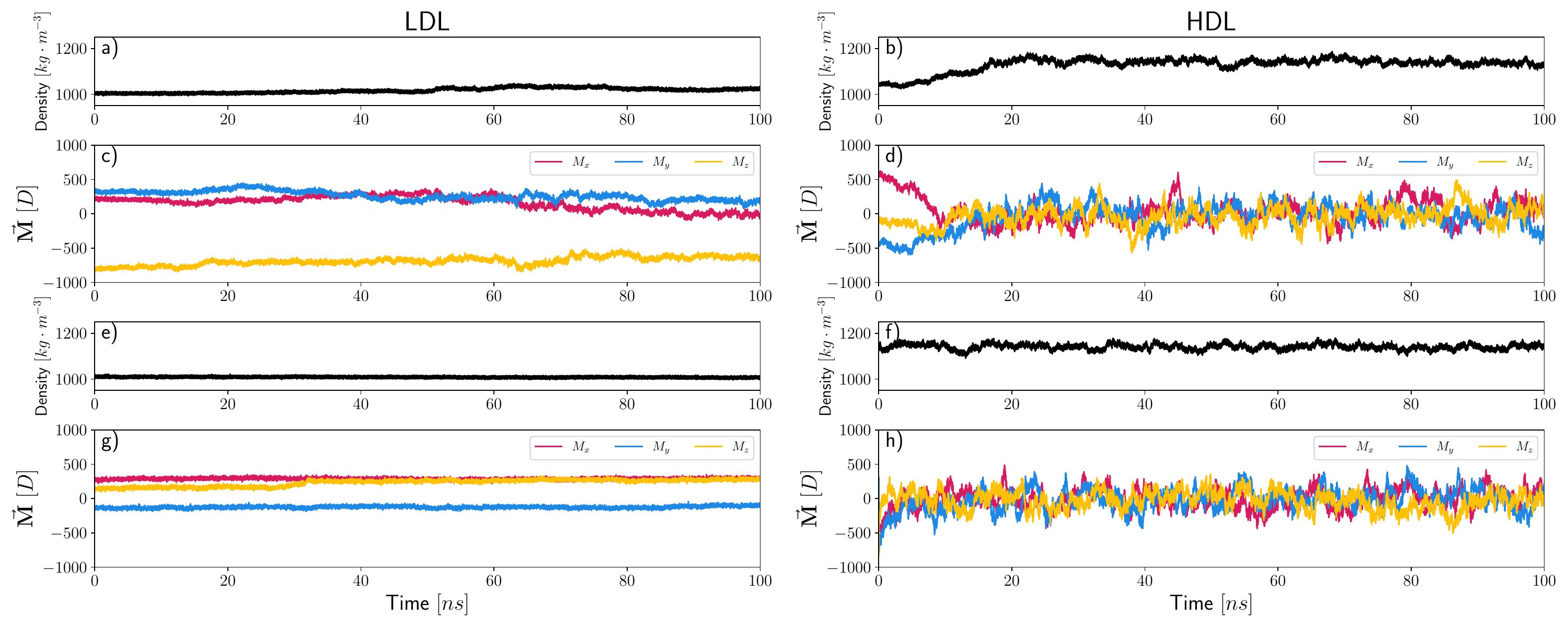}}
\caption{Time series of density and dipole components from four independent DNN simulations with 1536 water molecules at 235K and 3200 bar. Panels a, b, e and f show the density as a function of time for the different simulations. Panels c and g display dipole components for the LDL phase, with spontaneous polarization and collective dipole reorientations visible as a kink in g. Panels d and h show dipole components for the HDL phase, with dipole fluctuations around zero, indicating the absence of spontaneous polarization, and a transition from LDL to HDL in d.}
     \label{fig:big_dipole}
 \end{figure*}
The preceding analyses showing important polarization effects within the LDL phase as well as the coupling of density fluctuations to collective reorientational dynamics, are all inferred from rather small boxes consisting of 192 water molecules. Since orientational correlations may extend beyond 1nm, we wanted to assess the sensitivity of some of our key findings to finite size effects. It should be noted that substantial and long-lived polarization fluctuations are  also observed in the classical-potential (TIP4P/2005) simulations which feature a larger number of waters. To estimate the reliability of our observations, we conducted DNN simulations of systems consisting of 1536 water molecules at the same pT conditions of 235K and 3200 bar respectively. We maintained temperature and pressure constant using the Nose–Hoover barostat \cite{Nos1984,Hoover1985,Hoover1986}, setting the thermostat and the barostat relaxation times to 50fs and 500fs, respectively. For computational purposes, we employed a compressed version of the SCAN neural-network model \cite{Zhang2021}, utilizing the scheme developed in Ref. \cite{Lu2022}. The cut-off for this potential is 6\AA. We used a hydrogen mass of 2 AMU for the time integration, allowing us to employ a time step size of 0.5fs. Since these are significantly more expensive, we conducted four independent simulations, each of length 100ns of which 2 remain within the LDL phase, one transforms from LDL to HDL and the last remains in the HDL phase throughout. The four trajectories are illustrated in Figure \ref{fig:big_dipole}. 

The left panels display the density followed by the three dipole components for the LDL trajectories while the right ones are those for the HDL. Similar to the situation observed with the small boxes, the large samples also display a tendency to spontaneous polarization - the dipole components take on different magnitudes for the entire 100ns. In Figure \ref{fig:big_dipole}g, there is also an event similar to the kinks observed in the small box corresponding to collective dipole reorientations. Figure \ref{fig:big_dipole}d, instead, illustrates a trajectory where the spontaneous polarization irreversibly moves to the HDL phase where all components fluctuate around zero, a feature that is observed for the entire 100ns in Figure \ref{fig:big_dipole}h. Within the limits of the simulations currently available, we do not observe any striking differences in the magnitude of the polarization fluctuations between LDL and HDL going from the small to large box sizes.
\section*{Discussion and Perspectives}

In this work, we have taken advantage of state-of-the-art deep neural-network potentials which have recently confirmed the existence of a second critical point in deeply supercooled water. Harnessing these trajectories, we have revisited the putative liquid-liquid second critical point deep in the supercooled regime with specific focus on understanding the role of polarization in this transition.

Our analysis gives rise to a rather intriguing phenomena namely that the LDL phase presents strong propensity to display spontaneous polarization. This behavior is observed both in the structure along the dipole coordinates as well as in the behavior of the dynamics. By employing a recent data-driven approach that automatically detects large angular jumps, we show that the transitions between HDL and LDL appear to be triggered by an accumulation of large reorientational motions in the network.

These findings offer new insights and perspectives on behavior of supercooled water close to its critical point. On one side, it is clear that the kinetics within the LDL phase and transitions to the HDL are strongly modulated not only by the mass density order parameter but also by the polarization. This finding should motivate the design of new experimental probes looking specifically at dielectric anomalies under supercooled conditions, for example using applied external fields \cite{Cassone2024} which may alter the relative stability of the two phases. It is important to clarify that all our results point to the presence of non-zero polarization within the LDL phase, which we refer to as \textit{ferroelectric features}. However, this does not imply that the LDL phase is necessarily ferroelectric. An examination of the probability distribution along both the density and polarization (Figure 2) shows that the LDL phase displays some underlying structure which may manifest as a paraelectric glassy phase. Indeed, our data are compatible with a paraelectric phase with large dielectric susceptibility due to large polarization fluctuations. The observed transitions between different polarization states suggest a finite free-energy barrier, but it remains unclear whether this is due to intrinsic paraelectricity or finite-size effects. Additionally, the free-energy landscape might be rough with many minima, or the different observed dipole values may be due to finite-domain effects. Our analysis of different trajectories from larger samples indicates that further work involving enhanced sampling methods is necessary to reach more quantitative conclusions about the nature of the thermodynamics in the LDL polarized phase. There have been several previous theoretical efforts for example, by Men’shikov and Fedichev \cite{Menshikov2011}, and Pastore \textit{et al.} \cite{Izzo2024} to rationalize the thermodynamics associated with the coupling between the density and polarization.

Besides this, our report presents the first quantitative numerical demonstration of this effect from first principles simulations, offering new challenges from the computational perspective. Specifically, our results are limited to only sub-microsecond timescales and one would need to deploy enhanced sampling techniques along the density and polarization coordinates to adequately make stronger claims about thermodynamic and dynamical quantities. Finally, from a theoretical and perhaps conceptual side, the dynamical features we observed in the LDL phase are akin to what one observes in glassy systems under supercooling \cite{Biroli2011,Berthier2007_1,Berthier2007_2} and it remains an open question to the bearing of our results on the glass-transition of liquid water.

\section*{Materials and Methods}

\subsection*{Molecular dynamics simulations}

The trajectories used for the analysis taken from the reference \cite{Gartner2022}, consist of two different NPT simulations of 192 water molecules. The target temperature and pressure are set to 235K and 3200bar respectively and the two trajectories have been initialized with different and independent frames. The system dynamics are propagated by a DNN accurately trained on SCAN-DFT \cite{SCANPerdew} data. The training has been performed with DeePMD-kit \cite{deepmdv2}, while the MD simulations are performed with LAMMPS software \cite{Thompson2022}.  More details on the simulation conditions and the training procedure can be found in reference \cite{Gartner2022}.

\subsection*{Dipole Deep Neural Network}

To build the dataset for the training of the dipole DNN model, we collected configurations from different SCAN-DFT simulations of supercooled water performed at different densities and temperatures, ranging from undercooled to ambient conditions. Furthermore, we added frames directly taken from the trajectories of the reference \cite{Gartner2022}, to include specific snapshots of low- and high-density water. We then performed \textit{ab initio} calculations with the \qe distribution \cite{QE-2009,QE-2017,QE-2020}, using the SCAN \cite{SCANPerdew} functional, the plane-wave pseudopotential method, Hamann–Schluter– Chiang–Vanderbilt norm-conserving pseudopotentials \cite{Hamann2013} with a kinetic-energy cutoff of 150Ry. From the DFT electronic density, we computed the Maximally Localized Wannier Functions (MLWF) \cite{wannier,Souza2001} using the \texttt{wannier90} code \cite{Pizzi2020}. The centers of MLWF are called maximally localized Wannier centers (WC), and one can define the dipole of a water molecule as
\begin{equation*}
    \vec{\bm{\mu}} = +6e \ \vec{\bm{r}}_{O} +e \ \vec{\bm{r}}_{H_1} +e \  \vec{\bm{r}}_{H_2} -2e \sum_{i=1}^4 \vec{\bm{w}}_{i}
\end{equation*}
where $\vec{\bm{r}}_{O}$, $\vec{\bm{r}}_{H_1}$ and $\vec{\bm{r}}_{H_2}$ are the position of the oxygen and hydrogen atoms. Ultimately, $\vec{\bm{w}}_{i}$ refer to the positions of the four WC of the water molecule, one for each pair of valence electrons. Finally, the computed molecular dipoles are used to train the dipole DNN starting from the atomic positions.

The dipole DNN is constructed and trained with the DeePMD-kit \cite{deepmdv2}. The cutoff radius of the model is set to 6 \AA. The size of the embedding and fitting nets are (25, 50, 100) and (100, 100, 100), respectively. The training dataset is made of $\sim$5000 water configurations and the model was trained with 2 million steps of stochastic gradient descent.
 
\subsection*{Angular Jump Analysis and Coordination Defects}

In a recent work, some of us developed an approach that uses the time series of the components of the dipole moment of each water molecule and automatically identifies times when there are transitions corresponding to either small or large angular jumps and more generally reorientational dynamics. For more details, the reader is referred to the original Reference\cite{OffeiDanso2023}. Here we just summarise the essential ingredients behind the procedure.

The analysis relies on using the body fixed dipole vector $v(t)$ which is passed through a filtering procedure (see details of filtering protocol in the Supporting Information) to generate another time series ${\bf{v}}_{F}(t)$.
We subsequently proceed to calculate the derivative of ${\bf{v}}_{F}(t)$ using a finite difference approach. Subsequently, we perform a cross product between this derivative vector and ${\bf{v}}_{F}(t)$ to generate a new vector. This new vector corresponds to the direction perpendicular to the plane in which the body-fixed vector ${\bf{v}}_{F}(t)$ rotates 
\begin{equation}
    {{{{{{{\bf{n}}}}}}}}(t)={{{{{{{{\bf{v}}}}}}}}}_{{{{{\mbox{F}}}}}}(t)\times \frac{d{{{{{{{{\bf{v}}}}}}}}}_{{{{{\mbox{F}}}} }}(t)}{dt}.
\end{equation}
Our methodology defines an angular swing as a process that maintains the plane of rotation of the body-fixed vector ${\bf{v}}_{F}(t)$ unchanged. In other words, during one angular swing, the direction of ${\bf{n}}(t)$ remains constant. The start and end of swing events are then determined by significant instantaneous alterations in the direction of ${\bf{n}}(t)$. Specifically, we examine the following quantity 
\begin{equation}
    q(t)=1-\frac{{{{{{{{\bf{n}}}}}}}}(t)\cdot {{{{{{{\bf{n}}}}}}}}(t+dt)}{|{{{{{{{\bf{n}}}}}}}}(t)||{{{{{{{\bf{n}}}}}}}}(t+dt)|}.
\end{equation}
During the swing, this quantity remains zero. At the start and at the end of the swing, however, this quantity exhibits a non-zero value, indicating a shift in the plane of rotation of the body-fixed vector ${\bf{v}}_{F}(t)$. As a result, the initiation and termination times of angular swings can be determined by observing peaks in  ${\bf{q}}(t)$, which align with extremities in the filtered time series, ${\bf{v}}_{F}(t)$, effectively identifying sudden angular changes. From this analysis, we can essentially accumulate the statistics on the number of angular jumps observed involving angular jumps larger than a certain magnitude 
$\Delta\theta$.

The number of defects was determined by counting the total number of water molecules that are not hydrogen bonded. A geometrical criterion was used which considers two water molecules to be hydrogen bonded when the distance between the donating and accepting oxygen atoms (O$_D$ and O$_A$) is within 3.0 \AA \ and the angle formed by the bond vector between the donating hydrogen and oxygen (H$_D$ and O$_D$) and the bond vector between O$_D$ and O$_A$ is less than 60$^\circ$. Note that these criteria are different from those used in liquid water at room temperature because of the different thermodynamic conditions being simulated. In the SI (Figure 3) the radial (O$_D$-O$_A$ distance) and angular distribution functions of the hydrogen bond angles are shown. Although the absolute number of defects can change depending on the specific thresholds used, the trends are robustly reproduced.

\subsection*{Data Availability}
The data related to this work, including trained DNN model and the relative dataset, are publicly available for download at \url{https://github.com/cesaremalosso/PNAS-2024-ferroelectric_supercooled_water}
\subsection*{Acknowledgement}
This work was partially supported by the European Commission through the MaX Centre of Excellence for supercomputing applications (grant number 101093374) and by the Italian MUR, through the PRIN project ARES (grant number 2022W2BPCK) and the Italian National Centre from HPC, Big Data, and Quantum Computing (grant number CN00000013), founded within the Next Generation EU initiative. A.H acknowledges the funding
received by the European Research Council (ERC) under the European Union’s Horizon 2020 research and innovation programme (grant number 101043272 - HyBOP). C.M and S.B thank Roberto Car for insightful discussions. All the authors thank Francesco Sciortino for sharing the trajectories for the classical simulations.

\subsection*{Author contributions}
C.M., M.G.I., S.B., and A.H. conceive the research; C.M., S.B., and A.H performed research; C.M. analyzed data; N.M. contributed new analysis tools; C.M., S.B., and A.H. wrote the paper.
\subsection*{Competing Interests}
No conflict of interests is declared by any of the authors.
\printbibliography

\newpage
\title{SUPPLEMENTARY INFORMATION }
\maketitle
\section{Results for TIP4P/2005 simulations}

\captionsetup[subfigure]{labelformat=simple}
\renewcommand{\thesubfigure}{\alph{subfigure}}

\begin{figure}
\captionsetup[subfigure]{font=Large,labelfont=Large,labelformat=simple}
\begin{subfigure}{0.5\linewidth}
\centering
\subfloat[]{\includegraphics[width=\linewidth]{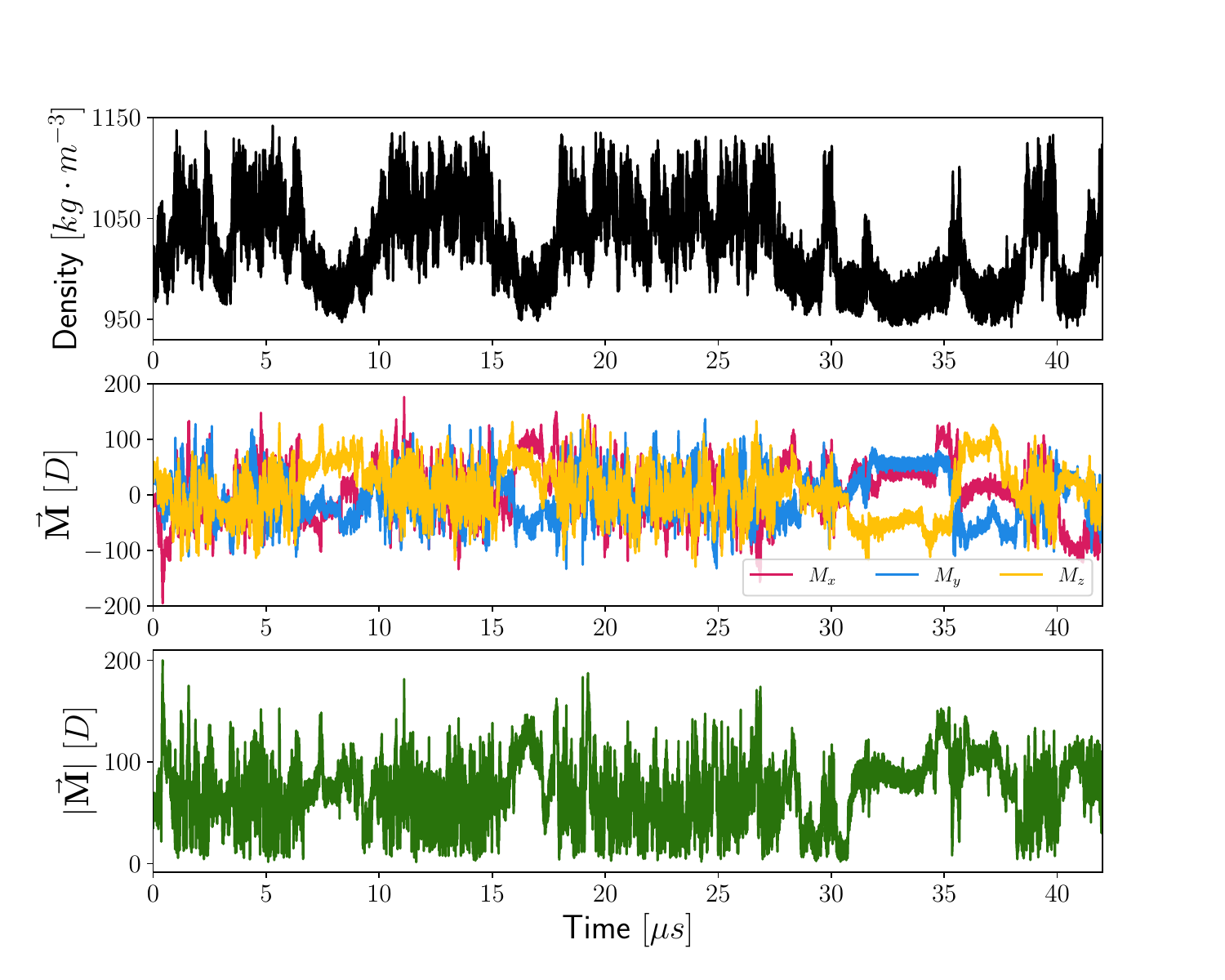}}
\end{subfigure}%
\begin{subfigure}{0.5\linewidth}
\centering
\subfloat[]{\includegraphics[width=\linewidth]{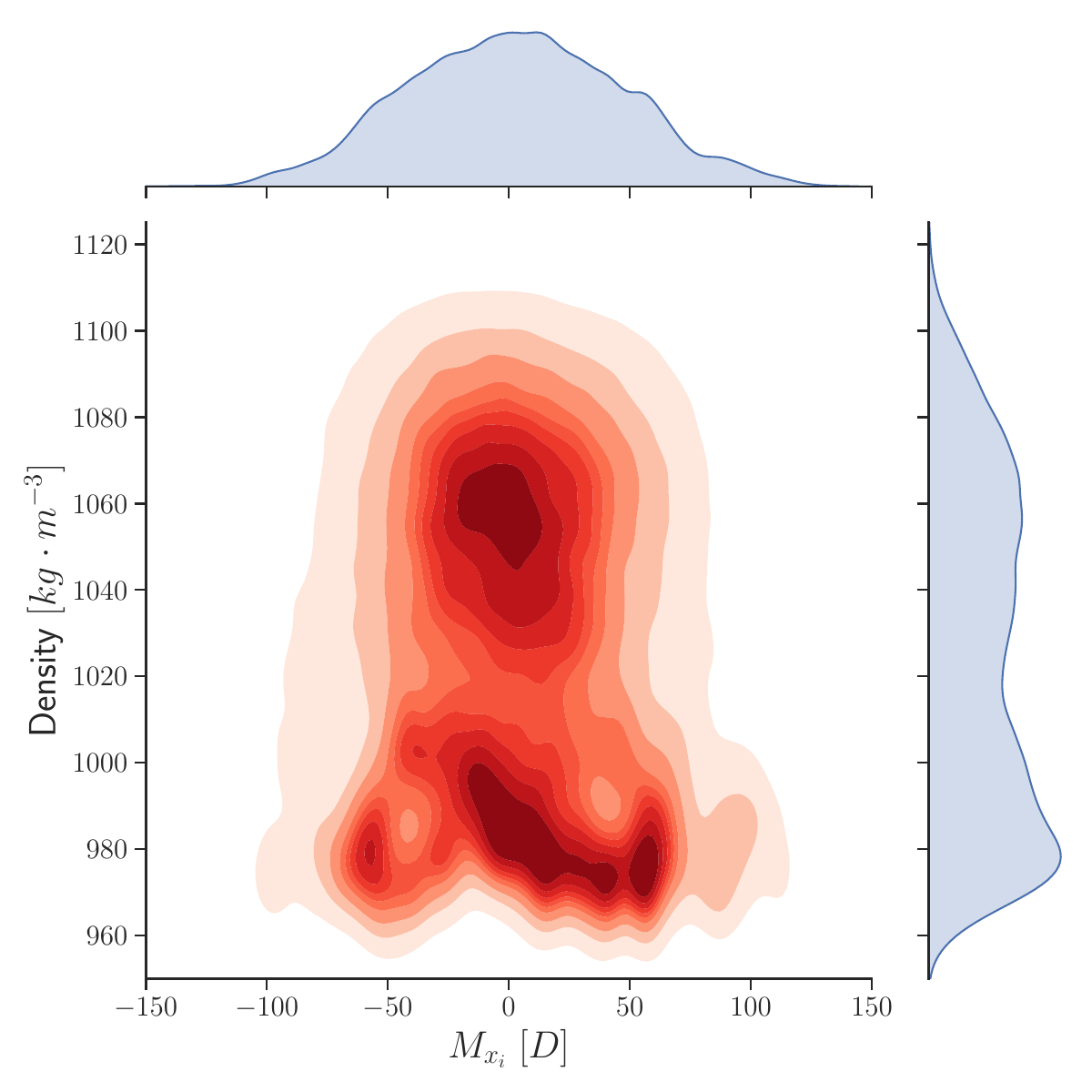}}
\end{subfigure}%
\caption{\textit{(a)}: Density[\textit{top}], total dipole[\textit{middle}] and its modulus[\textit{bottom}] fluctuations during isobaric-isothermal molecular dynamics simulations of TIP4P/2005 at nearly supercritical conditions - 177K and 1750bar. \textit{(b):} Statistics of critical fluctuations displayed as two-dimensional density and dipole probability densities.}
     \label{fig:dipole1}
 \end{figure}

We analyzed the coupling between the density and dipole over a microsecond-long TIP4P/2005 water simulation made of 300 molecules at nearly supercritical conditions - 177K
and 1750bar \cite{Debenedetti2020}. The dipole has been computed from the atomic coordinates and the point charges built into the model. The result of this analysis is shown in Fig. \ref{fig:dipole1}a. The different subplots show the temporal evolution of the density of the system (\textit{top}), the time series of each component of the total dipole (\textit{middle}), and finally of its modulus (\textit{bottom}). In Figure \ref{fig:dipole1}b, the density and the three equivalent components of the dipole are illustrated through a two-dimensional probability distribution. 

\begin{figure}[t]
\captionsetup[subfigure]{font=Large,labelfont=Large}
\begin{subfigure}{0.5\linewidth}
\centering
\subfloat[]{\includegraphics[width=\linewidth]{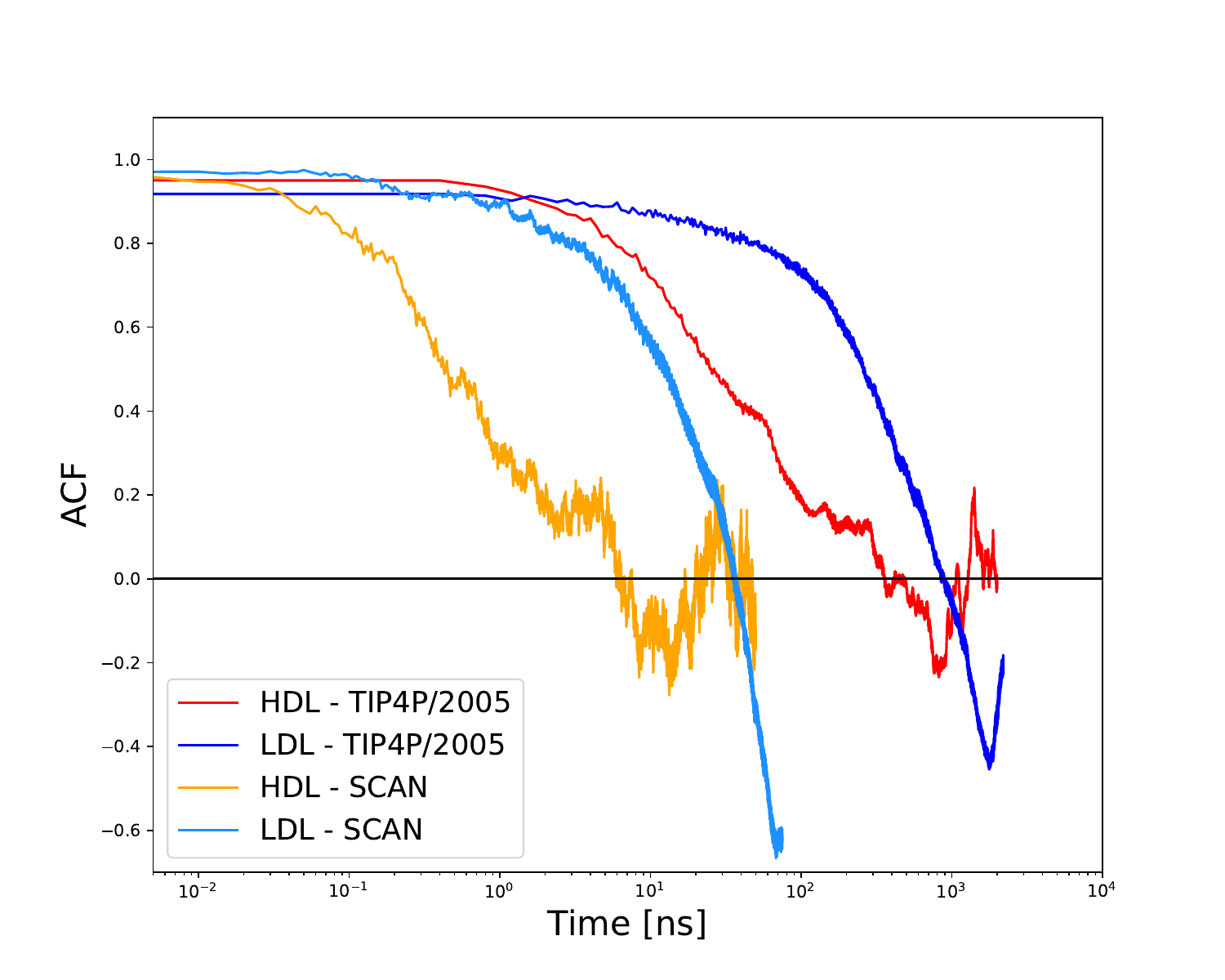}}
\end{subfigure}%
\begin{subfigure}{0.5\linewidth}
\centering
\subfloat[]{\includegraphics[width=\linewidth]{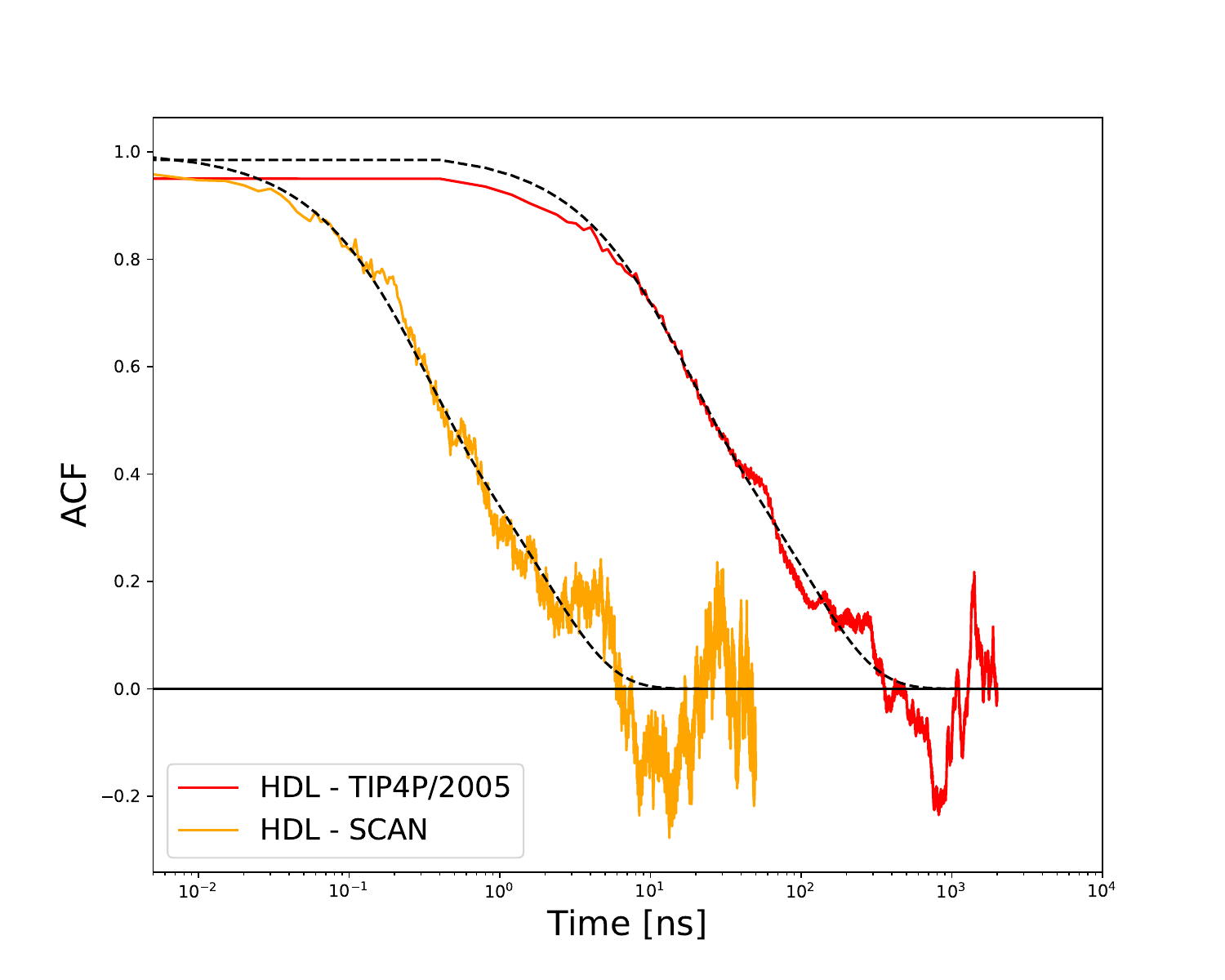}}
\end{subfigure}%
\caption{Left panel shows the total dipole correlation function computed for both TIP4P/2005 and SCAN-DFT models within low- and high-density supercooled water. Right panel shows the exponential fit to the dipole correlation of the HDL phase for the two models.}
     \label{fig:acf}
 \end{figure}

In Fig. \ref{fig:acf} we report the total dipole auto-correlation function (ACF) for both the TIP4P/2005 and the SCAN-DFT simulations \cite{Gartner2022}. The ACF is computed as:
\begin{equation*}
    ACF(t) = \frac{\langle (\vec{\textbf{M}}(t) - \langle \vec{\textbf{M}} \rangle)(\vec{\textbf{M}}(0) - \langle \vec{\textbf{M}} \rangle)\rangle}{\langle \vec{\textbf{M}}^2 \rangle - \langle \vec{\textbf{M}} \rangle^2}
\end{equation*}

For the HDL phase, the relaxation times associated with the dipole for SCAN and TIP4P/2005 model are ~2ns and ~150ns respectively, obtained by extracting the slow component associated with a bi-exponential fit shown in Fig \ref{fig:acf}b.

\begin{figure}[h]
\centering
\includegraphics[height=8cm]{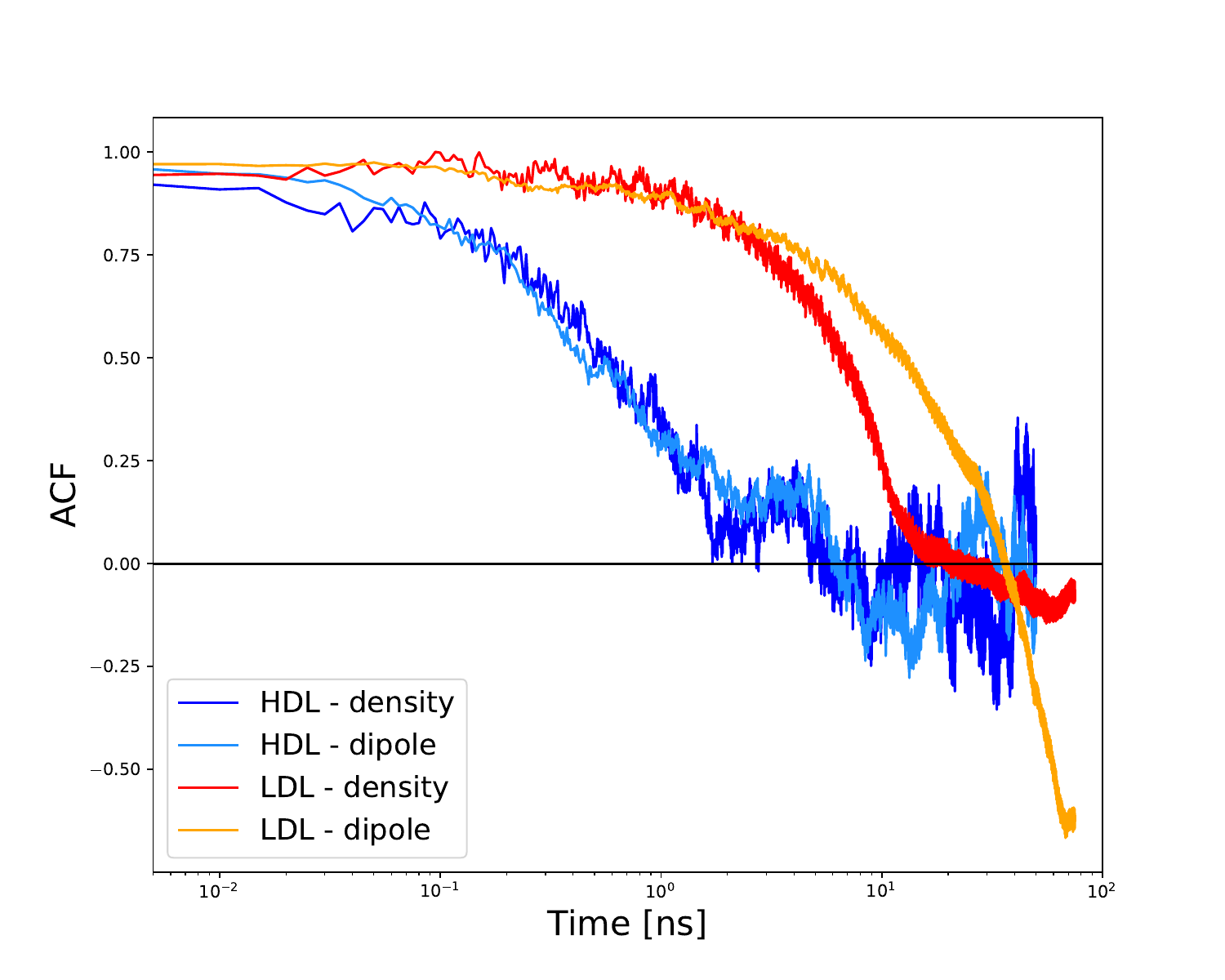}
\caption{Comparison between the total dipole time correlation function and the density time correlation function of the SCAN water model for both the HDL and LDL phases.}
\label{fig:acf_dipden}
\end{figure}

In Fig. \ref{fig:acf_dipden} we report the comparison between the time correlation of the total dipole and of the density within both the HDL and LDL phases of the SCAN-DFT model. While for HDL phase, density and polarization relax on similar timescales, in the LDL, the dynamics is significantly slower.

\section{Filtering Time Series}

To reduce the noise arising from high-frequency fluctuations in the dipole components, we applied a second-order low-pass digital butterworth filter (implemented using the SciPy.signal package in Python), engineered to maintain a near-flat frequency response within the passband. Sampling of the trajectories occurred every 5 picoseconds (ps), with a cutoff frequency set at 10GHz (1/100 in inverse ps), corresponding to the frequency at which the magnitude response of the filter reaches its threshold. In contrast to the room temperature, relaxation dynamics is much slower and hence one does not need a small time-step for processing the trajectories.

In addition to this initial filtering step aimed at reducing noise and preserving essential signal components, we further refined the trajectories by applying a mean filter using the \textit{smooth} function with a span of 20 data points (equivalent to a time window of 100 ps). This additional smoothing procedure facilitated the subsequent automated detection of angular swings by ensuring smoother trajectories.

Moreover, during the post-processing phase of the defect fractions and the count of swings within the time series, we introduced an extra layer of filtering using a low-pass butterworth filter with a cutoff frequency of 0.05GHz (1/20000 in inverse ps). This supplementary filtering step enabled us to effectively capture fluctuations occurring on the nanosecond (ns) scale. 

\begin{figure}
\captionsetup[subfigure]{font=Large,labelfont=Large}
\begin{subfigure}{0.5\linewidth}
\centering
\subfloat[]{\includegraphics[width=\linewidth]{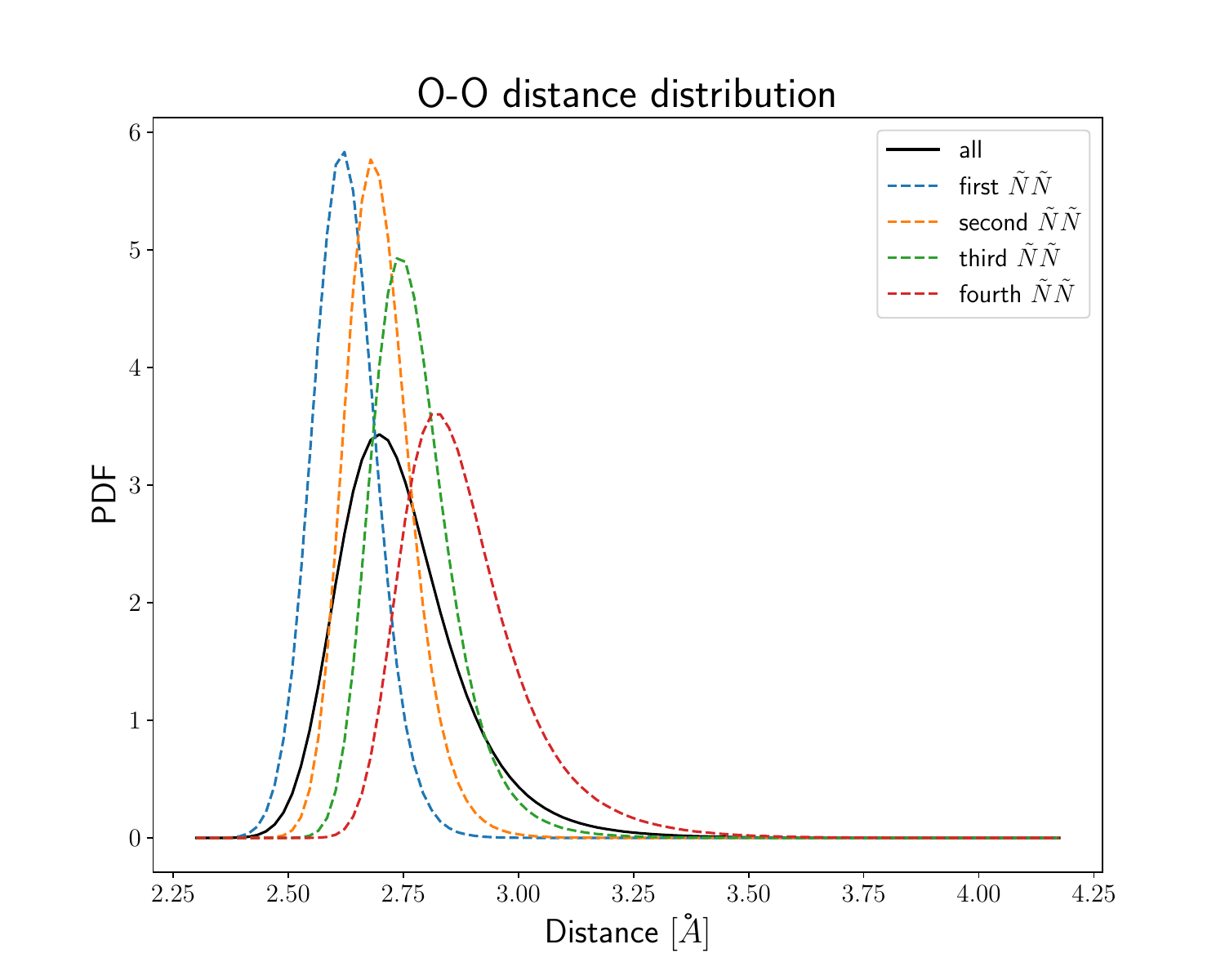}}
\end{subfigure}%
\begin{subfigure}{0.5\linewidth}
\centering
\subfloat[]{\includegraphics[width=\linewidth]{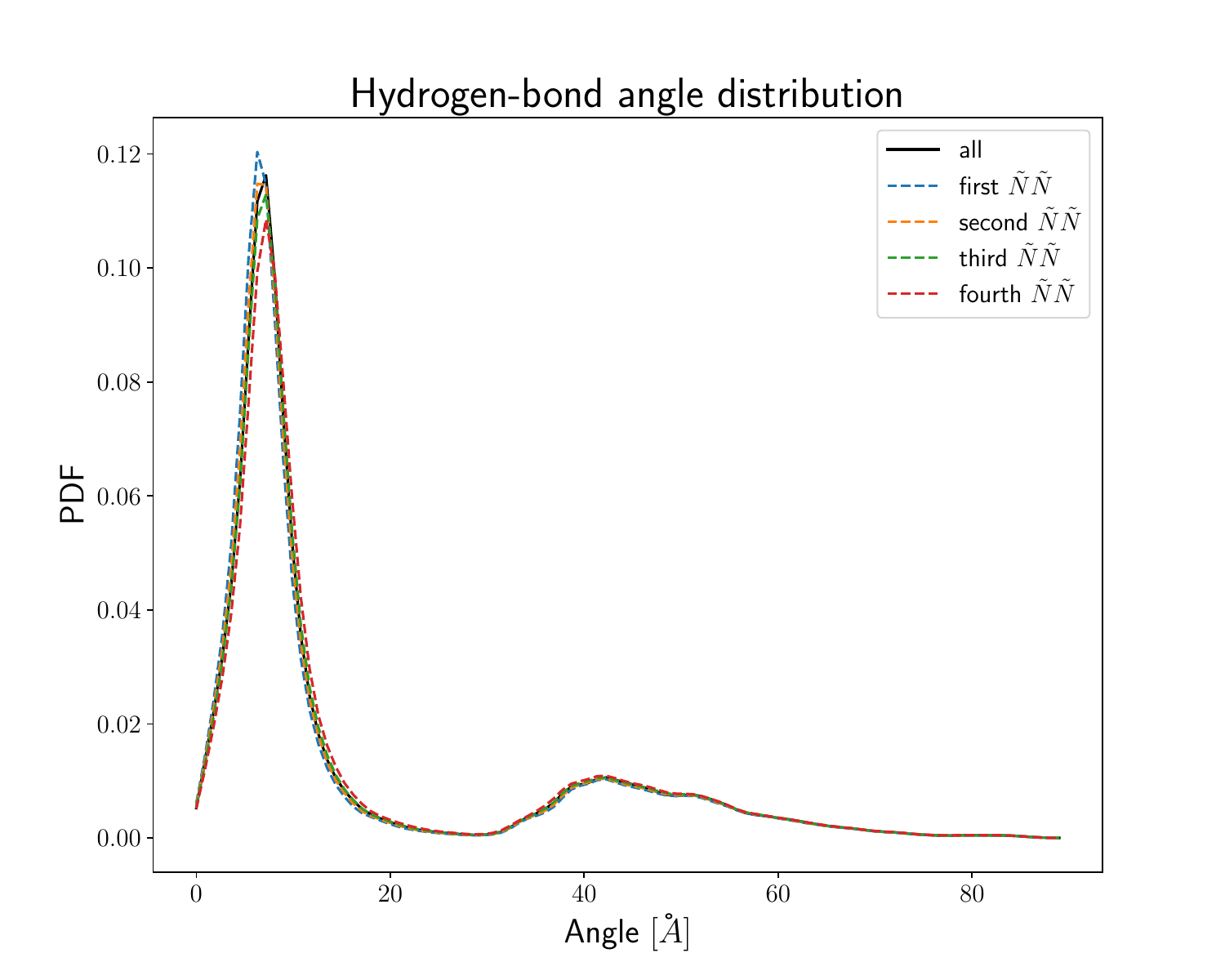}}
\end{subfigure}%
\caption{Left panel shows the oxygen-oxygen distance distribution function for the nearest 4 neighbours to every water molecule. Right panel shows the hydrogen-bond angle distribution for these 4 nearest neighbour waters.}
     \label{}
 \end{figure}

\end{document}